\documentclass[preprintnumbers,amsmath,amssymb,nofootinbib]{revtex4}
\usepackage{graphicx}
\usepackage{dcolumn}
\usepackage{bm}
\usepackage[dvips]{color}

\newcommand{\beq}{\begin{equation}}
\newcommand{\eeq}{\end{equation}}
\newcommand{\bq}{\begin{equation}}
\newcommand{\eq}{\end{equation}}
\newcommand{\ba}{\begin{array}}
\newcommand{\ea}{\end{array}}
\newcommand{\beqa}{\begin{eqnarray}}
\newcommand{\eeqa}{\end{eqnarray}}

\def\[{\left[}
\def\]{\right]}
\def\({\left(}
\def\){\right)}

\evensidemargin=\oddsidemargin

\def\pslash{\not{\hbox{\kern-4pt $p$}}}
\def\qslash{\not{\hbox{\kern-4pt $q$}}}
\def\lv{\not{\hbox{\kern-4pt $L$}}}
\def\lsim{\mathrel{\raise.3ex\hbox{$<$\kern-.75em\lower1ex\hbox{$\sim$}}}}
\def\gsim{\mathrel{\raise.3ex\hbox{$>$\kern-.75em\lower1ex\hbox{$\sim$}}}}
\def\ifmath#1{\relax\ifmmode #1\else $#1$\fi}

\begin{document}

 \title{$\boldsymbol{W_L W_L}$ Scattering in Higgsless Models: Identifying Better Effective Theories}
 
\author{ Alexander S. Belyaev$^a$, R.\ Sekhar Chivukula$^b$, Neil D. Christensen$^b$,  \\
Hong-Jian He$^c$, Masafumi Kurachi$^d$, Elizabeth H.\ Simmons$^b$, and Masaharu Tanabashi$^e$}

\affiliation{$^a$ School of Physics \& Astronomy, University of Southampton,
Highfield, Southampton SO17 1BJ, UK\\
Particle Physics Department, Rutherford Appleton Laboratory, Chilton,
Didcot, Oxon OX11 0QX, UK\\
$^b$\ Department of Physics and Astronomy, Michigan State University, 
East Lansing, MI 48824, USA\\
$^c$\ Center for High Energy Physics, Tsinghua University, 
Beijing 100084, China\\
$^d$\ Theoretical division T-2, Los Alamos National Laboratory, Los Alamos, NM 85745, USA\\
$^e$\ Department of Physics, Nagoya University, 
Nagoya 464-8602, Japan}

\date{\today}

\preprint{SHEP-09-15} 
\preprint{MSUHEP-090715} 
\preprint{LA-UR-09-01080} 
\preprint{NTLP 2009-01}

\begin{abstract}

The three-site model has been offered as a benchmark for studying the collider phenomenology of Higgsless models. In this paper we analyze how well the three-site model performs as a general exemplar of Higgsless models in describing $W_L W_L$ scattering, and which modifications can make it more representative.  We employ general sum rules relating the masses and couplings of the Kaluza-Klein (KK) modes of the gauge fields in continuum and deconstructed Higgsless models as a way to compare the different theories.  We show that the size of the four-point vertex for the (unphysical) Nambu-Goldstone modes and the degree to which the sum rules are saturated by contributions from the lowest-lying KK resonances both provide good measures of the extent to which a highly-deconstructed theory can accurately describe the low-energy physics of a continuum 5d Higgsless model.  After comparing the three-site model to flat and warped continuum models, we analyze extensions of the three-site model to a longer open linear moose with an additional $U(1)$ group and to a ring (``BESS' or ``hidden local symmetry") model with three sites and three links. Both cases may be readily analyzed in the framework of the general  sum rules.  We demonstrate that $W_LW_L$ scattering in the ring model can very closely approximate scattering in the continuum models, provided that the hidden local symmetry parameter `$a$'  is chosen to mimic $\rho$-meson dominance of $\pi\pi$ scattering in QCD.  The hadron and lepton collider phenomenology of both extended models is briefly discussed, with a focus on the complementary information to be gained from precision measurements of the $Z'$ line shape and $ZWW$ coupling at a high-energy lepton collider. 

\end{abstract}

\date{\today}
 
 \maketitle

\section{Introduction}

The origin of electroweak symmetry breaking continues to be a question of great interest as we enter the LHC era.  While the Standard Model (SM) can produce electroweak symmetry breaking through introduction of a scalar Higgs doublet, the lack of experimental evidence for this approach makes it advisable to consider alternatives such as Higgsless models \cite{Csaki:2003dt,Csaki:2003zu} that can accommodate electroweak 
symmetry breaking without the introduction of elementary scalars \cite{Higgs:1964ia}.  In these theories, which are
based on compactified five-dimensional gauge theories, the scattering of longitudinal electroweak gauge bosons ($W_L$ and $Z_L$) is unitarized via the exchange of massive vector bosons 
\cite{SekharChivukula:2001hz,Chivukula:2002ej,Chivukula:2003kq},
the Kaluza-Klein (KK) modes of the theory.  Moreover, under the rubric of AdS/CFT duality \cite{Maldacena:1997re}, Higgsless models may be thought of as ``dual" to models of dynamical electroweak symmetry
breaking, such as walking \cite{Holdom:1981rm,Holdom:1984sk,Yamawaki:1985zg,Appelquist:1986an}
technicolor \cite{Weinberg:1979bn,Susskind:1978ms}.  

A leading challenge for non-Standard models of electroweak symmetry breaking lies in the strong constraints placed on new physics by precision electroweak observables \cite{Peskin:1992sw,Altarelli:1990zd,Altarelli:1991fk,Barbieri:2004qk}.  The phenomenology of the most general Higgsless model may be studied \cite{Foadi:2003xa,Hirn:2004ze,Casalbuoni:2004id,Chivukula:2004pk,Perelstein:2004sc,Georgi:2004iy,SekharChivukula:2004mu} by using the technique of deconstruction \cite{ArkaniHamed:2001ca,Hill:2000mu} and computing the precision electroweak parameters \cite{Chivukula:2004af} in a related linear moose model \cite{Georgi:1985hf}. 
In general one finds that the KK vector-bosons responsible for unitarizing $W_LW_L$ scattering in Higgsless models, which are the analogs of the technivector mesons in technicolor models, mix with the electroweak bosons and alter their properties with respect to Standard Model predictions. However, Higgsless models that incorporate ideally \cite{SekharChivukula:2005xm} delocalized fermions \cite{Cacciapaglia:2004rb,Foadi:2004ps}, in which the ordinary fermions propagate appropriately in the compactified extra dimension (or in deconstructed language, derive their weak properties from more than one site on the lattice of gauge groups \cite{Chivukula:2005bn,Casalbuoni:2005rs}), yield phenomenologically acceptable values for all $Z$-pole observables. In this case the leading deviations from the Standard Model appear in multi-gauge-boson couplings, rather than the $S$ and $T$ parameters.  

These issues have previously been explored  \cite{SekharChivukula:2006cg,Matsuzaki:2006wn,Sekhar Chivukula:2007ic,Abe:2008hb,Abe:2009ni} in the context of a highly deconstructed Higgsless model  in which the infinite Kaluza-Klein tower is collapsed to only one electroweak triplet of extra vector bosons; that is, the electroweak gauge group has the form $SU(2) \times SU(2)\times U(1)$.  The gauge-sector of this ``three-site'' model \cite{SekharChivukula:2006cg} is a particular 
BESS model of extended  electroweak gauge symmetries \cite{Casalbuoni:1985kq,Casalbuoni:1996qt}  motivated by models of hidden local symmetry \cite{Bando:1985ej,Bando:1985rf,Bando:1988ym,Bando:1988br,Harada:2003jx}.  It reproduces the most prominent low-energy effects of a continuum Higgsless model quite accurately, yet remains simple enough that it can be encoded in a Matrix Element Generator program for use with Monte Carlo simulations.  Initial studies of the LHC phenomenology of the three-site model, focusing on $W'$ boson production in the vector-boson fusion and associated production modes, were reported in \cite{He:2007ge}, production of nearly-fermiophobic vector bosons in the three-site model was described in \cite{Ohl:2008ri}, and a more extensive analysis including additional channels is underway \cite{He2009}.  Related studies of other Higgsless scenarios, both continuum and deconstructed, are also in the literature \cite{Birkedal:2004au,Accomando:2008dm,Accomando:2008jh,Alves:2008up,Englert:2008wp,Hirn:2006wg,Hirn:2007bb,Hirn:2008tc}. 

The three-site model has been offered as a benchmark or test case for studying $W_L W_L$ scattering in  Higgsless models.   It is therefore appropriate to consider how well the three-site model performs as a general representative of Higgsless models: how can one quantify the differences between the three-site model and continuum models? how closely is it related to other deconstructed models? and, how universal is its phenomenology?  These are the questions this paper will address.

We will employ the sum rules \cite{Csaki:2003dt, SekharChivukula:2008mj}  relating the masses and couplings of the KK modes in continuum and deconstructed models as a way to compare the different theories.  These identities enable us to quantify how well a given theory performs at unitarizing the scattering of electroweak gauge bosons at a particular energy scale, a key element of any theory of electroweak symmetry breaking.  We will see that the tendency of the sum rules to be saturated \cite{Birkedal:2004au} by contributions from the lowest-lying KK resonances provides a good measure of the extent to which a highly-deconstructed theory like the three-site model can accurately describe the low-energy physics.  In addition, one must take into account the fact that deconstructed and continuum Higgsless models are generally distinguished by the respective presence and absence of a four-point vertex for the scattering of (unphysical) Nambu-Goldstone modes; this provides a valuable point of comparison.

In section II, we set the stage by looking at these sum rules in the context of a pair of continuum models: $SU(2) \times SU(2)$ Higgsless models in either a flat or warped five-dimensional space.   Since these two models have previously been discussed in the literature, we will remind the reader of the key properties needed to calculate the quantities relevant to our comparisons, and will refer the reader to Ref. \cite{Chivukula:2005ji} for further details.  Comparing the results in the two continuum models to one another and to the analogous quantities in the three-site model 
(which we also briefly describe) will provide the template for studying how well any given deconstructed model does as an effective theory of a continuum Higgsless model.  

We then consider whether there are modifications of the three-site model that can remedy some of its shortcomings in reproducing the results of continuum models, as measured by the sum rules.  Ref. \cite{SekharChivukula:2008mj} discussed how to extend the relevant sum rules to mooses with extra $U(1)$ groups and to ring models, and we will study both kinds of scenarios explicitly in this paper.  In section III, we introduce a slightly longer\footnote{We will term this model the ``mimimal'' four-site model to distinguish it from the $SU(2)^3 \times U(1)$ models studied elsewhere \cite{Chivukula:2005bn,Accomando:2008dm,Accomando:2008jh,SekharChivukula:2008gz}.}  open linear moose \cite{Georgi:1985hf} based on the group $SU(2)\times SU(2) \times U(1) \times U(1)$; in section IV, we introduce a model that has three sites and also a third link field closing the moose to form a triangular ring, a general
BESS or hidden local symmetry model \cite{Bando:1985ej,Bando:1985rf,Bando:1988ym,Bando:1988br,Harada:2003jx,Casalbuoni:1985kq,Casalbuoni:1996qt}.  In each case we will show the key features of the model in the main text (with further details in an appendix) and then focus on comparing the model to the three-site model and continuum theories.  We find that the following quantity, $\hat{a}$, which is based on the four-point coupling of the Nambu-Goldstone bosons 
\begin{equation}
 \hat{a} \equiv \frac43 (1 - g_{\pi\pi\pi\pi}) \approx \frac{v^2 g_{Z' WW}^2 M_{Z'}^2}{M_W^4}~,
 \label{eq:firstahat}
 \end{equation}
provides a useful measure of the ability of the deconstructed models to approximate the behavior of the continuum theories.  To the extent that the lightest neutral KK boson ($Z'$) saturates \cite{Birkedal:2004au} the sum rules, $\hat{a}$ is also related to the $Z'$ boson's mass and coupling to $W$ pairs, as on the far-right of Eq.~(\ref{eq:firstahat}) where $v = (\sqrt{2} G_F)^{-1/2} \approx 250 $ GeV.  In continuum theories, $\hat a = 4/3$ and the multi-pion coupling $g_{\pi\pi\pi\pi}$ vanishes; in the linear deconstructed models we study, $\hat a = 1$ and $g_{\pi\pi\pi\pi} \neq 0$; in the triangular ring model, $\hat{a}$ is a free parameter equal to the parameter $a$ of hidden local symmetry (HLS) models \cite{Bando:1985ej,Bando:1985rf,Bando:1988ym,Bando:1988br,Harada:2003jx}, allowing interpolation between these cases.   For this reason $W_LW_L$ scattering in the ring model can, if the hidden local symmetry $a$ parameter is chosen to equal four-thirds,  very closely approximate scattering in the continuum models.  Note that the value $a=4/3$ corresponds in HLS models of QCD to $\rho$-meson dominance of $\pi\pi$
scattering \cite{Harada:2003jx,Harada:2006di}. On the other hand, when $a=1$ it is possible to demonstrate an equivalence between the ring model and an $[SU(2)]^3\times U(1)$ linear model, as we discuss in section V.

Section VI discusses briefly  the hadron and lepton collider phenomenology of the extended models.  We will see that LHC searches for $W'$ bosons in the vector boson fusion or associated production channels (with $W'$ decaying to $WZ$) tend to give a universal signature for all of the models discussed here.  Likewise, the expected contributions of the neutral KK states to $WW$ scattering are very similar in all the models discussed here; in the case of the four-site model with an extra $U(1)$ group, this is because we find there is a relationship between the neutral boson couplings to $W$ pairs in the four-site and three-site models:
\begin{equation}
 \left(  g_{Z'WW}^2 + g_{Z''WW}^2\right)_{\rm four-site} = \left( g_{Z'WW}^2 \right)_{\rm three-site}  .
\end{equation}
In contrast, the bounds on these models  arising  from LEP-II measurements of the $ZWW$ vertex provide complementary information that can distinguish among models; the same would be true for future measurements of the $Z'$ line shape or the $ZWW$ coupling at a linear electron-positron collider.

Our conclusions are presented in section VII.

\section{Comparison using $W_L W_L \to W_LW_L$ sum rules}

In both continuum and deconstructed Higgsless models, the gauge boson KK modes unitarize \cite{SekharChivukula:2001hz,Chivukula:2002ej,Chivukula:2003kq} the scattering of longitudinal vector bosons by cutting off the tendency of the scattering amplitudes to grow with powers of the scattering energy, $E$.  In order to play this role effectively, the KK mode couplings and masses must obey a set of identities derived in  \cite{Csaki:2003dt,SekharChivukula:2008mj}.  Hence, our discussion of the three-site model, its extensions, and related continuum models  will focus on the triple-gauge-boson couplings involving one KK mode and two light modes, $g_{Z'WW}$ and $g_{ZW'W}$; these contribute directly to KK exchange diagrams involved in longitudinal vector boson scattering processes. 

\subsection{Continuum Sum Rules and Triple Gauge Couplings}

In any continuum five-dimensional gauge theory, the sum rules that guarantee the absence, respectively, of ${\cal{O}}(E^4)$ and ${\cal{O}}(E^2)$ growth in the amplitude for $W^+_L W^-_L \rightarrow W^+_L W^-_L$ elastic scattering have the following form \cite{Csaki:2003dt,SekharChivukula:2008mj},
\begin{equation}
 \sum_{i=1}^{\infty} g_{Z_{i}WW}^2 = g_{WWWW} - g_{ZWW}^2 - g_{\gamma WW}^2 \ ,
\label{eq:E4sum}
\end{equation}
\begin{equation}
 3 \sum_{i=1}^{\infty} g_{Z_{i}WW}^2M_{Z_{i}}^2
 = 4g_{WWWW}M_W^2  
 - 3g_{ZWW}^2M_Z^2 \ ,
\label{eq:E2sum}
\end{equation}
where $Z_i$ represents the $i$-th KK mode of the neutral gauge boson. ($Z_1$ is also called $Z'$ and $Z_2$ is also called $Z''$.)  These are special cases [$n=m=0$] of the general identities given in  \cite{SekharChivukula:2008mj} for $W_{Ln} W_{Ln} \to W_{Lm} W_{Lm}$ scattering processes where two KK modes of level $n$ scatter into two modes of level $m$.

We will start by examining these identities in the context of an $SU(2)_A\times SU(2)_B$ gauge theory with either a flat or a warped extra dimension, with ideally delocalized fermions.   Most relevant details of these models (the action, mode expansion of the gauge bosons, boundary conditions, etc.) are given in subsections 2.1 and 2.2 of \cite{Chivukula:2005ji}.   However, before evaluating the sum rules in these models, we must calculate the  heavy-light-light triple gauge couplings $g_{Z'WW}$ and $g_{ZW'W}$, which were not computed in that paper.

\subsubsection{$SU(2)\times SU(2)$ Higgsless model with a flat extra dimension}
\label{sec:twoa}

The first model we consider is a 5d continuum $SU(2)_A \times SU(2)_B$ gauge theory in which the 5th dimension is flat.  As described in  \cite{Chivukula:2005ji}, the coordinate of the flat extra dimension may be expressed as the dimensionful coordinate $z$ which has boundaries at $0$ and $\pi R$; where convenient, we also employ the related dimensionless coordinate $\tilde{z}\equiv z/(\pi R)$. We examine the case where the $SU(2)_A$ and $SU(2)_B$ 
gauge bosons have the same bulk coupling, i.e. $g_{5A}=g_{5B}\equiv g_{5}$.  The boundary conditions at $z=0$ explicitly break $SU(2)_A \times SU(2)_B$ down to $SU(2)_W \times U(1)_Y$, where we identify $SU(2)_W$ with $SU(2)_A$ and hypercharge with the $T^3$ component of  $SU(2)_B$.  We also introduce $SU(2)_W$ and $U(1)_Y$ kinetic terms on the $z=0$ brane and  call the associated couplings $g_0$ and $g_Y$, respectively.  It is useful to define an angle $\theta$ relating these couplings as:
\begin{equation}
\frac{g_Y}{g_0} = \frac{\sin\theta}{\cos\theta} \equiv \frac{s_0}{c_0} 
\equiv t_0, 
\label{eq:t0}
\end{equation}
with $s^2_0 + c^2_0=1$, and also a parameter $\lambda$ relating $g_0$ to the bulk coupling
\begin{equation}
 \lambda \equiv (\pi R)\frac{g_0^2}{g_{5A}^2+g_{5B}^2}
 = (\pi R)\frac{g_0^2}{2 g_{5}^2}\ \  \left( \ll 1 \right).
\label{eq:lambda}
\end{equation}
We derive here the masses and wavefunctions for the $W_1$ and $Z_1$ bosons; those 
for the light $W$ and $Z$  can be found in \cite{Chivukula:2005ji}.  Note that in order to obtain the correct leading-order expressions for the triple gauge couplings, we need to keep the next-to-leading terms in the expressions for the wavefunctions.

Starting from the differential equations and boundary conditions in \cite{Chivukula:2005ji}, we find that the lowest-lying charged heavy gauge boson, $W_1 \equiv W'$, has a mass
\begin{equation}
 \pi R M_{W'} = 
  \frac{\pi}{2} + \frac{2}{\pi}\lambda \ +\ O(\lambda^2) ,\
\end{equation}
with a corresponding wavefunction
\begin{eqnarray}
 \chi_{W'}^A(z) &=& C_{W'}
    \left[ \sin(M_{W'}z) - \frac{4\lambda}{\pi}\cos(M_{W'}z)
    \right]~,
\\
 \chi_{W'}^B(z) &=& C_{W'}\ \sin(M_{W'}z)~,
\end{eqnarray}
whose normalization factor is
\begin{equation}
  C_{W'} = \frac{g_0}{\sqrt{2\lambda}}
    \left[ 
        1-\frac{4 \lambda}{\pi^2}           
      + O(\lambda^2)
    \right]~.
\end{equation}
The lowest-lying neutral heavy gauge boson, $Z_1 \equiv Z'$, likewise, has a mass
\begin{equation}
 \pi R M_{Z'} = 
  \frac{\pi}{2} + \frac{2}{c^2_0 \pi}\lambda \ +\ O(\lambda^2) ,\
\end{equation}
with a corresponding wavefunction
\begin{eqnarray}
 \chi_{Z'}^A(z) &=& C_{Z'}
    \left[ \sin(M_{Z'}z) - \frac{4\lambda}{\pi}\cos(M_{Z'}z)
    \right]~,
\\
 \chi_{Z'}^B(z) &=& C_{Z'}
    \left[ \sin(M_{Z'}z) - t_0^2\frac{4\lambda}{\pi}\cos(M_{Z'}z)
    \right]~,
\end{eqnarray}
and normalization factor
\begin{equation}
  C_{Z'} = \frac{g_0}{\sqrt{2\lambda}}
    \left[ 
        1-\frac{4 \lambda}{c_0^2\pi^2}           
      + O(\lambda^2)
    \right]~.
\end{equation}
Note that the $Z'$ becomes degenerate with the $W'$ in the limit    $c_0 \rightarrow 1$ ($s_0, t_0 \rightarrow 0$), in which the boundary conditions for $\chi_{Z'}^B(z)$ at $z=0$ are of Dirichlet form, like those for $\chi_{W'}^B(z)$.  In addition, in the limit $\lambda \rightarrow 0$, the wave functions of the $W'$ and $Z'$ are expressible in terms of a pure sine function. This is consistent with the fact that  taking $\lambda \rightarrow 0$ corresponds to imposing Dirichlet boundary conditions on both the $SU(2)_A$ and $SU(2)_B$ gauge fields at $z=0$.

The triple gauge boson couplings that involve one heavy gauge boson can be calculated 
as\footnote{The more compact notation c or s is used in these expressions where
the difference between employing the bare and corrected weak
angles would cause only higher-order corrections.} :
\begin{eqnarray}
 g_{Z'WW} &=& 
 \int_0^{\pi R}dz
  \frac{1}{g_{5}^2} \left\{ \chi_{Z'}^A(z)
   \left|\chi_{W}^A(z)\right|^2 +  
          \chi_{Z'}^B(z)
	  \left|\chi_{W}^B(z)\right|^2 
 \right\}  + \frac{1}{g_{0}^2} \chi_{Z'}^A(0)
   \left|\chi_{W}^A(0)\right|^2
 \nonumber \\
 &=& -g_0 \frac{8\sqrt{2}}{\pi^3} \lambda^{1/2} + O(\lambda^{3/2}) \nonumber \\
   &=& - \frac{4\sqrt{2}}{\pi^2}\,\frac{e}{s} \left(\frac{M_W}{M_{W'}}\right) +   O\left[\left(\frac{M_W}{M_{W'}}\right)^3\right],
\label{eq:Z'WW_flat}
\end{eqnarray}
\begin{eqnarray}
 g_{ZW'W} &=& 
 \int_0^{\pi R}dz
  \frac{1}{g_{5}^2}
  \left\{  
    \chi_{Z}^A(z)
    \chi_{W'}^A(z)\chi_{W}^A(z) 
  \right. +
  \left.
          \chi_{Z}^B(z)
	  \chi_{W'}^B(z)\chi_{W}^B(z)
  \right\} 
  + \frac{1}{g_{0}^2} \chi_{Z}^A(0)
   \chi_{W'}^A(0)\chi_{W}^A(0)
 \nonumber \\
&=&
-\frac{g_0}{c}\,\frac{8\sqrt{2}}{\pi^3} \lambda^{1/2} + O(\lambda^{3/2})
 \nonumber \\
   &=& - \frac{4\sqrt{2}}{\pi^2}\,\frac{e}{sc} \left(\frac{M_W}{M_{W'}}\right) + 
   O\left[\left(\frac{M_W}{M_{W'}}\right)^3\right].
\label{eq:ZW'W_flat}
\end{eqnarray}
We will use these results to compare models a little later in this paper.  The masses, wavefunctions, and triple gauge couplings of the higher KK gauge boson modes may be calculated in similar fashion, but will not be required for our purposes.

\subsubsection{$SU(2)\times SU(2)$ Higgsless model with a warped extra dimension}

The next model we consider is a 5d continuum $SU(2)_A \times SU(2)_B$ gauge theory with a warped 5th dimension.
As described in  \cite{Chivukula:2005ji}, we adopt the conformally flat metric
$
ds^2 = \left( \frac{R}{z} \right)^2 
\left( \eta_{\mu \nu} dx^\mu dx^\nu - dz^2 \right),
$
for the ${\rm AdS_5}$ space, and require that the coordinate $z$ be 
restricted to the interval 
$
R \left(\equiv e^{-b/2} R'\right)\le z \le  R'~ 
$.
We assume a large hierarchy between the ``Planck'' brane at $z=R$ and the ``TeV'' brane at $z = R'$
(equivalent to taking $b \gg 1$). 
We take the $SU(2)_A$ and $SU(2)_B$ 
gauge bosons to have the same bulk coupling, i.e. 
$g_{5A}=g_{5B}\equiv g_{5}$.  In order to arrange for a non-trivial weak mixing angle, we introduce a brane-localized $U(1)_Y$ kinetic term on the Planck brane, where $U(1)_Y$ corresponds to the $T^3$ component of $SU(2)_B$; the dimensionless coupling associated with this kinetic term is $g_Y$.   Then the electromagnetic coupling $e$ is related to $g_Y$ and $g_5$ 
\begin{equation}
 \frac{1}{e^2} =
  \frac{bR}{g_{5}^2}+
  \frac{1}{g_Y^2}.
\end{equation}
We can identify quantities $c_0$, $s_0$ that satisfy the relation
$s_0^2 + c_0^2 = 1$ and may be interpreted as the cosine and sine 
of the bare weak mixing angle:
\begin{equation}
 \frac{c_0^2}{e^2} = \left(\frac{bR}{2}\right) \frac{1}{g_{5}^2} 
  + \frac{1}{g_Y^2}~,\ \ \ \ 
 \frac{s_0^2}{e^2} = \left(\frac{bR}{2}\right) \frac{1}{g_{5}^2}~.
\end{equation}
Because expressions for the masses and wavefunctions of the light $W$ and $Z$ bosons 
can be found in \cite{Chivukula:2005ji},  we show only those for the $W'$ and $Z'$  below.

Starting from the differential equations and boundary conditions in  \cite{Chivukula:2005ji}, we find that the $W'$ boson has a mass
\begin{equation}
 R' M_{W'} = x_1 + \frac{\pi}{2}\frac{Y_0(x_1)}{J_1(x_1)}\left( \frac{1}{b}\right) + O\left(\frac{1}{b^2}\right),
\end{equation}
where $J_n(z)$ and $Y_n(z)$ represent the $n$-th order Bessel functions of the first and second kind respectively, and $x_1$ is the first zero of $J_0(z)~$, i.e. $J_0(x_1)=0$.
The corresponding wavefunction of $W'$ is 
\begin{eqnarray}
 \chi_{W'}^A(z) &=& C_{W'}
    \left[ z J_1(M_{W'}z) +\left(\frac{\pi}{b}\right) z Y_1(M_{W'}z) + 
\cdots  \right],
\\
 \chi_{W'}^B(z) &=& C_{W'} z J_1(M_{W'}z) \left[1+\left(\frac{\pi}{b}\right)\frac{Y_1(x_1)}{J_1(x_1)}+ \cdots\ \right] ~,
\end{eqnarray}
where $C_{W'}$ is a normalization constant.
In the expressions above, ellipses represent higher order 
terms with respect to $1/b$.  The $Z'$ mass may be written as
\begin{equation}
 R' M_{Z'} = x_1 + \frac{\pi}{2 c^2_0}\frac{Y_0(x_1)}{J_1(x_1)}\left( \frac{1}{b}\right) + O\left(\frac{1}{b^2}\right),
\end{equation}
with a corresponding wavefunction
\begin{eqnarray}
 \chi_{Z'}^A(z) &=& C_{Z'}
    \left[ z J_1(M_{Z'}z) +\left(\frac{\pi}{b}\right) z Y_1(M_{Z'}z) +\ 
\cdots\  \right] ,
\\
 \chi_{Z'}^B(z) &=& C_{Z'}
   \,\left[ z J_1(M_{Z'}z) +\left(\frac{\pi}{b}\right) \frac{s^2_0}{c^2_0} z Y_1(M_{Z'}z)+
\cdots  \right]\,  \left[1+\left(\frac{\pi}{b}\right)\frac{c^2_0-s^2_0}{c^2_0}\frac{Y_1(x_1)}{J_1(x_1)} + \cdots \right] ~ ,
\end{eqnarray}
where $C_{Z'}$ is a normalization constant. 

Since the calculations of the normalization constants $C_{W', Z'}$ and the 
triple gauge couplings involve integrations whose results are 
not expressible in simple closed form, we carry out the integrations numerically. 
The resultant values of $g_{Z'WW}$ and $g_{ZW'W}$ are, at leading order, proportional 
to the input value of $M_W/M_{W'}$ and can be written as 
\begin{eqnarray}
 g_{Z'WW} &=& 
 \int_{R}^{R'}dz \left(\frac{R}{z}\right)
  \frac{1}{g_{5}^2} \left\{ \chi_{Z'}^A(z)
   \left|\chi_{W}^A(z)\right|^2 
+
          \chi_{Z'}^B(z)
	  \left|\chi_{W}^B(z)\right|^2 
 \right\} 
\label{eq:Z'WW_warped_integral}
\nonumber\\
 &\simeq& -~0.36\ \left(\frac{M_W}{M_{W'}}\right)
  + O\left[\left(\frac{M_W}{M_{W'}}\right)^3\right]\,,
\label{eq:Z'WW_warped}
\end{eqnarray}

\begin{eqnarray}
 g_{ZW'W} &=& 
 \int_{R}^{R'}dz \left(\frac{R}{z}\right)
  \frac{1}{g_{5}^2}\left\{  \chi_{Z}^A(z)
    \chi_{W'}^A(z)\chi_{W}^A(z) 
+
          \chi_{Z}^B(z)
	  \chi_{W'}^B(z)\chi_{W}^B(z)
 \right\} ,
\label{eq:ZW'W_warped_integral}
\nonumber\\
&\simeq&
 -~0.41\ \left(\frac{M_W}{M_{W'}}\right)
  + O\left[\left(\frac{M_W}{M_{W'}}\right)^3\right].
\label{eq:ZW'W_warped}
\end{eqnarray}
We will use these results to compare models in the next section.  Again, the masses, wavefunctions, and triple gauge couplings of higher KK modes may  be calculated by similar methods, but will not be needed for our purposes.

\subsection{Comparisons and implications}

One way to compare models is to look simply at the values of the heavy-light-light triple gauge couplings.  Doing this for 
 the flat (Eqs. (\ref{eq:Z'WW_flat}) and  (\ref{eq:ZW'W_flat})) and warped  (Eqs. (\ref{eq:Z'WW_warped})  and (\ref{eq:ZW'W_warped})) continuum $SU(2)_A\times SU(2)_B$ Higgsless models yields an interesting result.   Making the numerical approximations $\frac{4\sqrt{2}}{\pi^2}\frac{e}{s}\simeq 0.36$ 
and $\frac{4\sqrt{2}}{\pi^2}\frac{e}{s c}\simeq 0.41$ in Eqs. (\ref{eq:Z'WW_flat}) and  (\ref{eq:ZW'W_flat}), we find,
\begin{equation}
\frac{g_{Z'WW}|_{\rm warped-5d}}{g_{Z'WW}|_{\rm flat-5d}} \simeq 
\frac{g_{ZW'W}|_{\rm warped-5d}}{g_{ZW'W}|_{\rm flat-5d}} \simeq 1 .
\label{eq:similarities}
\end{equation}
In other words, the values of $g_{Z'WW}$ and $g_{ZW'W}$ in these continuum models are essentially independent of the 5d geometry to leading order.   This contrasts with the situation for the couplings among the light bosons.   Because the first KK mode nearly saturates the LHS of (\ref{eq:E4sum}) while the value of $g_{\gamma WW}$ is set by gauge invariance, the difference $g_{WWWW} - g^2_{ZWW}$ is independent of geometry; however, the individual couplings,  $g_{ZWW}$ and $g_{WWWW}$, are different for different continuum geometries \cite{Chivukula:2005ji}.

If we obtain the analogous couplings in the three-site model \cite{SekharChivukula:2006cg} by accounting for the wavefunction overlap among the gauge boson mass eigenstates at each site:
\begin{eqnarray}
g_{Z'WW} &=& g\, v^0_{Z'}(v^0_W)^2  + \tilde{g}\, v^1_{Z'}(v^1_W)^2 =  -\,\frac{e}{2 s_Z} \left(\frac{M_W}{M_{W'}}\right)  + \cdots \,,
\label{eq:Z'WW_3-site}\\
g_{ZW'W} &=& g\, v^0_{Z}v^0_{W'}v^0_W + \tilde{g}\, v^1_{Z}v^1_{W'}v^1_W =  -\,\frac{e}{2 s_Z c_Z} \left(\frac{M_W}{M_{W'}}\right) + \cdots \,,
\label{eq:ZW'W_3-site}
\end{eqnarray}
where the $Z$-standard weak mixing angle is defined as
\begin{equation}
 s_Z^2 c_Z^2 \equiv \frac{e^2}{4\sqrt{2} G_F M_Z^2}~,
 \end{equation}
then we can compare the continuum model triple gauge couplings to those in the three-site
 model, assuming a common value for $M_W/M_{W'}$:
\begin{equation}
\frac{g_{Z'WW}|_{\rm three-site
}}{g_{Z'WW}|_{\rm flat-5d}} \simeq 
\frac{g_{ZW'W}|_{\rm three-site
}}{g_{ZW'W}|_{\rm flat-5d}} \simeq 
\frac{\pi^2}{8\sqrt{2}} \simeq 0.87 \,.
\end{equation}
The values of $g_{Z'WW}$ and $g_{ZW'W}$ in the three-site
 Higgsless model are about 13\% smaller than those values in 5-dimensional $SU(2)_A\times SU(2)_B$ Higgsless models.  

An alternative comparison focuses on the degree to which the first KK mode saturates the sum on the LHS of the identities (\ref{eq:E4sum}) and (\ref{eq:E2sum}).   Suppose that we form the ratio of the $n=1$ term in the sum on the LHS to the full combination of terms on the RHS, evaluated to leading order in $(M_W / M_{W_1})^2$.  The ratio derived from (\ref{eq:E4sum}) is
\begin{equation}
\frac{g_{Z'WW}^2}{g_{WWWW} - g_{ZWW}^2 - g_{\gamma WW}^2 } ~,
\label{eq:firstratio}
\end{equation}
while that derived from (\ref{eq:E2sum}) is
\begin{equation}
\frac{3 g_{Z'WW}^2M_{Z'}^2}{ 4g_{WWWW}M_W^2  - 3g_{ZWW}^2M_Z^2}~.
\label{eq:secondratio}
\end{equation}
If the $n=1$ KK mode saturates the identity, then the related ratio will be 1.0;  ratio values less than 1.0 reflect contributions from higher KK modes.  We see from Table 1 that each of these ratios is nearly 1.0 in both the $SU(2)_A \times SU(2)_B$ flat and warped Higgsless models \cite{Birkedal:2004au}, confirming that the first KK mode nearly saturates the sum rules in these continuum models.  The similar behavior of the two 5d models is consistent with our finding that the $g_{Z'WW}$ coupling is relatively independent of geometry.  

Because the first KK mode nearly saturates the identities (\ref{eq:E4sum}) and (\ref{eq:E2sum}) in these continuum models, the ratios (\ref{eq:firstratio}) and (\ref{eq:secondratio}) should be useful for drawing comparisons with the three-site model, which only possesses a single KK gauge mode.  As shown in the 3rd column of Table \ref{tab:1}, the first ratio has the value one in the three-site model, meaning that the identity  (\ref{eq:E4sum}) is still satisfied.  The ratio related to identity (\ref{eq:E2sum}), however, has the value 3/4 for the three-site
 model, meaning that the second identity is not satisfied; the longitudinal gauge boson scattering amplitude continues to grow as $E^2$ due to the underlying non-renormalizable interactions in the three-site model. Since the value of the denominator has not changed appreciably, this indicates a difference between the values of the $g_{Z'WW}$ couplings in the continuum and three-site models, as we discuss in the next section.

\begin{table}[hbt]
\begin{center}
\begin{tabular}{|c|c|c|c|}
\hline
&\  5d $2\times 2$ Flat\  &\  5d $2\times 2$ Warped\  &\  Three-site\ 
 \\ \hline &&&\\
$ \frac{g_{Z'WW}^2}{g_{WWWW} - g_{ZWW}^2 - g_{\gamma WW}^2 }$
& 
$\frac{960}{\pi^6} \simeq 0.999$
&
$0.992$
&
$1$ 
\\ [3mm] \hline &&& \\
$ \frac{3 g_{Z'WW}^2M_{Z'}^2
 }{ 4g_{WWWW}M_W^2  
 - 3g_{ZWW}^2M_Z^2}$
&
$\frac{96}{\pi^4} \simeq 0.986$
&
$0.986$
&
3/4
\\  [3mm] \hline
\end{tabular}
\end{center}
\caption{Ratios relevant to evaluating the degree of cancellation of growth in the $W_L W_L$ scattering amplitude from the lowest lying $KK$ resonance at order $E^4$ (top row, from Eq. (\protect\ref{eq:firstratio})), and at order $E^2$ (second row, from Eq. (\protect\ref{eq:secondratio})). A value close to one indicates a high degree of cancellation from the lowest lying resonance. Shown in successive columns for the $SU(2)_A \times SU(2)_B$ flat and warped continuum models discussed in the text, and the three-site deconstructed model.}
\label{tab:1}
\end{table}


\subsection{Deconstruction sum rules}

\begin{figure}[b]
\centering
\includegraphics[width=0.6\textwidth]{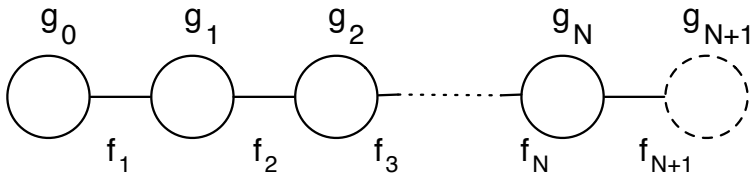}
\caption
{A general linear deconstructed Higgsless model \protect\cite{SekharChivukula:2008mj}, 
in moose notation \protect\cite{Georgi:1985hf}. The groups at sites 0 to $N$ are $SU(2)$ gauge
groups, and the group at site $N+1$ is a global $SU(2)$ group with a gauged $U(1)$ subgroup, and each link corresponds to an
$SU(2) \times SU(2)/SU(2)$ nonlinear sigma model breaking the adjacent groups down
to their diagonal sum. The coupling constants ($g_i$)
and $f$-constants ($f_j$) may be given values corresponding
 to the position-dependent coupling and warp factors of a related continuum model.
 The phenomenological constraint that $W$ and $Z$ have masses much less than the $KK$
vector resonances is equivalent (by using the appropriate block-spin transformation) to taking $g_0,\, g_{N+1} \ll g_1,\ldots,g_{N}$ \protect\cite{Sekhar Chivukula:2006we}.
The three-site model corresponds to $N=1$ and $f_1 = f_2=\sqrt{2} v$ 
\protect\cite{SekharChivukula:2006cg}.
}
\label{fig:generalmoose}
 \end{figure}

To understand the behavior of the two ratios (\ref{eq:firstratio},\ref{eq:secondratio}) in the three-site
 model, we need to relate them to the sum rules that apply in  a general deconstructed model illustrated in Fig. \ref{fig:generalmoose}; the gauge sector of the three-site model corresponds to setting $N=1$ and $f_1 = f_2 = \sqrt{2} v$ in Fig. \ref{fig:generalmoose}. As shown in \cite{SekharChivukula:2008mj}, the sum rule in Eq.~(\ref{eq:E4sum}), which is related to canceling $E^4$ growth in the longitudinal scattering amplitude, is the same in deconstructed models as in 5d models.  Hence, it is not surprising that the upper ratio in Table \ref{tab:1} has the same value in the three-site model as in the continuum models.  As noted above for the three-site model, however, in a 
 general deconstructed Higgsless model the $W_L W_L$ scattering
 amplitude continues to grow like $E^2$ at high-energies \cite{Chivukula:2002ej}. 
 This behavior is most easily understood in
 terms of the Equivalence Theorem
 \cite{Cornwall:1974km,Vayonakis:1976vz,Lee:1977eg,Chanowitz:1985hj,Yao:1988aj,Bagger:1990fc,He:1992ng,He:1993qa,He:1994yd,He:1997cm} which relates the $W_L W_L$ scattering amplitude
 to the scattering amplitude of the unphysical Nambu-Goldstone bosons that remain
 in a renormalizable gauge. The residual $E^2$ growth is related to the
 non-zero contact interactions among unphysical Nambu-Goldstone fields $\pi$
 present in deconstructed models,
and  the second sum rule  \cite {SekharChivukula:2008mj} takes on the modified 
form\footnote{The four-pion coupling $g_{\pi\pi\pi\pi}$ is equivalent to the dimensionless quantity $\tilde{G}_4^{0000}$ in Ref.~\cite{SekharChivukula:2008mj}.  Ref.~\cite{Da Rold:2005zs} instead uses the dimensionful quantity $g_{4\pi} \equiv g_{\pi\pi\pi\pi} / v^2$ to describe the four-pion vertex. } 
\begin{equation}
 4\frac{M_W^4}{v^2} g_{\pi\pi\pi\pi} + 3 \sum_{i} g_{Z_{i}WW}^2M_{Z_{i}}^2
 = 4g_{WWWW}M_W^2  
 - 3g_{ZWW}^2M_Z^2 \ .
\label{eq:E2sumd}
\end{equation}
Note this is an exact tree-level result in an arbitrary linear (or ring \cite {SekharChivukula:2008mj}) deconstructed Higgsless model, and that the $g_{\pi\pi\pi\pi}$ contact 
interactions vanish in the 5d continuum limit so that Eq.~(\ref{eq:E2sumd}) reduces
to (\ref{eq:E2sum}) for a continuum model.
Since the values of $g_{WWWW}$, $g_{ZWW}$, $M_W$ and $M_Z$ are very close to their Standard Model values in all of the models we are discussing, the right-hand sides of Eqs. (\ref{eq:E2sum}) and (\ref{eq:E2sumd}) are nearly identical.  The  presence of the non-zero $g_{\pi\pi\pi\pi}$ in (\ref{eq:E2sumd}) is what constrains $g_{Z'WW}$ to take on a significantly different value in the three-site model than it has in the continuum models.

Calculating the pion-pion scattering amplitude provides a related way of comparing the continuum 5d Higgsless models with the three-site
 model.  For simplicity, we frame this discussion in terms of a ``global'' theory; for the deconstructed model this corresponds to taking $g_0=g_{N+1}=0$ (which corresponds to setting the electroweak
 couplings, $g$ and $g'$, equal to 0), while 
in the case of continuum Higgsless models, this corresponds to imposing Dirichlet boundary conditions for all gauge fields at both ends of the interval.  Ref.~\cite{Sekhar Chivukula:2006we} studied $\pi\pi$ scattering in an arbitrary global linear moose and  obtained the following 
sum rule (a generalization of one originally derived by Da Rold and Pomarol 
\cite{Da Rold:2005zs}):
\begin{equation}
  \frac{1}{v^2} =  \frac{g_{\pi\pi\pi\pi}}{v^2} + 3 \sum_{i}\frac{g_{V_i\pi\pi}^2}{M_{V_i}^2}.
\label{eq:DPsum1}
\end{equation}
Here,  $g_{V_i\pi\pi}$ is the 
coupling\footnote{In the limit $g,\,g' \to 0$, one finds $g_{V_i \pi \pi} \to
\frac12 g_{Z_i W W} (M_{V_i}^2 / M_W^2)$; away from $g = g' = 0$, there will be corrections of order $M_W^2 / M_{W_i}^2$.} of $\pi\pi$ to the $i$-th vector boson KK-mode, and $M_{V_i}$ is the mass 
of the $n$-th KK-mode, $V_i^{a\mu}$.  In the continuum limit, the four-pion contact interaction 
vanishes, and Eq.~(\ref{eq:DPsum1}) reads 
\begin{equation}
  \frac{1}{v^2} = 3 \sum_{i}\frac{g_{V_i\pi\pi}^2}{M_{V_i}^2} , 
\label{eq:DPsum2}
\end{equation}
which should be compared with the celebrated KSRF relation
\cite{Kawarabayashi:1966kd,Riazuddin:1966sw}
for the $\rho$ meson in hadron dynamics (but note the different coefficients):
$1/f_\pi^2=2g_{\rho\pi\pi}^2/M_\rho^2$. 

In order to assess the ability of extensions of the three-site model to provide a better approximation to the low-energy behavior of continuum models, we define the parameter
\begin{equation}
\hat{a} \equiv \frac43 (1 - g_{\pi\pi\pi\pi})~.
\label{eq:secondahat}
\end{equation}
As noted earlier, $\hat{a} = 4/3$ in continuum models where $g_{\pi\pi\pi\pi} = 0$; in contrast, $\hat{a} = 1$ in the three-site model because $g_{\pi\pi\pi\pi} \approx 1/4$.   Relating this parameter to the 
sum rules above is enlightening.  Recalling that the values of $g_{WWWW}$ and $g_{ZWW}$ are, to lowest order in $M^2_W/M^2_{W_i}$, the same as their respective Standard Model values of $e^2/s_Z^2$ and $e c_Z / s_Z$, we find that the RHS of equation (\ref{eq:E2sumd}) takes the form
\begin{equation}
 4g_{WWWW}M_W^2  - 3g_{ZWW}^2M_Z^2 = \frac{4 M_W^4}{v^2} \,.
\end{equation}
This lets us use (\ref{eq:E2sumd}) to recast $\hat{a}$ as
\begin{equation}
 \hat{a} \equiv \frac43 (1 - g_{\pi\pi\pi\pi}) = \sum_{i} \frac{ v^2 g_{Z_{i}WW}^2M_{Z_{i}}^2}{M_W^4}\,.
 \label{eq:thirdahat}
\end{equation}
On the other hand, starting from the sum rule of Da Rold and Pomarol (\ref{eq:DPsum1}) allows us to make the connection
\begin{equation}
 \hat{a} \equiv \frac43 (1 - g_{\pi\pi\pi\pi}) = \sum_{i} \frac{4 v^2 g^2_{V_i \pi\pi}}{M_{V_i}^2}~.
 \label{eq;fourthahat}
\end{equation}
That the right-hand sides of Eqs. (\ref{eq:thirdahat}) and (\ref{eq;fourthahat}) are not only consistent, but actually equivalent term by term, may be seen by remembering that the Equivalence Theorem  \cite{Cornwall:1974km, Vayonakis:1976vz,Lee:1977eg, Chanowitz:1985hj, Yao:1988aj, Bagger:1990fc, He:1992ng,He:1993qa, He:1994yd, He:1997cm} and its extra-dimensional
counterpart  \cite{SekharChivukula:2001hz} show
that the $Z_iWW$ 
vertex with coupling strength $g_{Z_iWW}$ is equivalent at leading order to 
 the $V_i \pi\pi$ vertex with coupling $g_{V_i \pi\pi} = \frac12 g_{Z_iWW} (M^2_{V_i}/ M_W^2)$, where $\pi$ is the Nambu-Goldstone mode eaten by the $W$.  Finally, in both the three-site and continuum models, $\hat{a}$ takes on an even simpler form:
 \begin{equation}
 \hat{a} \approx \frac{ v^2 g_{Z_{1}WW}^2M_{Z_{1}}^2}{M_W^4} \approx \frac{4 v^2 g^2_{V_1 \pi\pi}}{M_{V_1}^2}\,.
\label{eq:fifthahat}
 \end{equation}
In the three-site model, the relationships in Eq. (\ref{eq:fifthahat}) are actually simple equalities because there is only a single $Z'$ boson.  In the case of continuum theories, the approximate relationships hold because the sums in the right-hand sides of Eqs. (\ref{eq:thirdahat}) and (\ref{eq;fourthahat})  are nearly saturated by the first terms \cite{Birkedal:2004au}. For Eq. (\ref{eq:thirdahat}) this relates to the previously-discussed saturation of (\ref{eq:E4sum}) and (\ref{eq:E2sum}) by the lowest KK mode; for Eq. (\ref{eq;fourthahat}), the results of Ref.~\cite{Sekhar Chivukula:2006we} imply that the RHS of (\ref{eq:DPsum2}) is dominated by the contribution of $V_1$ in both the flat and warped global $SU(2)\times SU(2)$ Higgsless models.

To test these ideas, let us look at an example. Since we expect that taking the number of sites in a linear deconstructed model to infinity should reproduce a continuous extra dimension, it is instructive to see what happens to the value of $\hat{a}$ in this limit.  Consider the linear moose in
Fig. \ref{fig:generalmoose} with $g_0=g_{N+1}=0$, and $N$ gauged $SU(2)$ groups in the simplest ``flat" case where all the decay constants are identical ($f_i = v \sqrt{N+1}$) and all the gauge couplings  $g_i$ are identical ($g_i = \sqrt{N}\tilde{g}$).
By diagonalizing the mass-squared matrix of the gauge bosons and its dual matrix, 
we can calculate the mass of $V_1$ and the wavefunctions of $\pi$ and $V_1$ from which 
we obtain the expression for $g_{V_1\pi\pi}$. The resultant expression for the 
parameter $\hat{a}$ takes the following form:
\begin{equation}
\hat{a} = \frac{2}{N(N+1)^3}\frac{\sin^2\frac{\pi}{N+1}}{\sin^6\frac{\pi}{N+1}}
    = \frac{128}{\pi^4}\left[1+\frac{1}{N}-\frac{\pi^2}{12N^2}+\cdots\right].
\end{equation}
Asymptotically, we find $\hat{a} \rightarrow 128/\pi^4 = 1.31405 \simeq 4/3$; equivalently one finds that  the value of $g_{\pi\pi\pi\pi}$ falls off as $1/N$, approaching zero at large $N$.  In other words, the values of $\hat{a}$ and $g_{\pi\pi\pi\pi}$ do approach their continuum counterparts, as anticipated.

\section{``Minimal'' Four-site Model with an additional $U(1)$ group}

\begin{figure}[tb]
  \begin{center}
    \includegraphics[height=4cm]{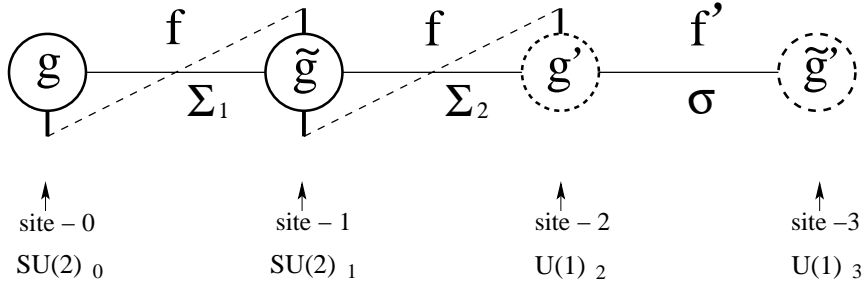}
  \end{center}
\caption{$SU(2)_0 \times SU(2)_1 \times U(1)_2 \times U(1)_3$ 
model analyzed in Section III. The solid circles represent (from left) $SU(2)_0$ and $SU(2)_1$
gauge groups, with coupling strengths $g_0 \equiv g$ and $g_1 \equiv {\tilde g}$, 
and dashed circles represent (from left) $U(1)_2$ and $U(1)_3$ 
gauge groups with couplings $g_2 \equiv g'$ and $g_3 \equiv {\tilde g}'$. 
The left-handed fermions, denoted by the lower vertical lines, are located at the sites 0 and 1, and the right-handed fermions, denoted by the upper vertical lines, at sites 1 and 2. 
The dashed lines correspond to Yukawa couplings, as described in Ref.~\cite{SekharChivukula:2006cg}. As discussed below, we will take $f_1=f_2=f\equiv \sqrt{2} v$ and take ${\tilde g}, {\tilde g}' \gg g, g'$. }
\label{fig:2-2}
\end{figure}

As mentioned in the Introduction, our aim is to seek modifications of the three-site model \cite{SekharChivukula:2006cg}  that can remedy some of its shortcomings, as measured by the deconstruction sum rules above.  Because Ref. \cite{SekharChivukula:2008mj} discussed how to extend the relevant sum rules to mooses with extra $U(1)$ groups, the first extension we consider is a four-site linear moose model that includes an extra $U(1)$ group; we term this the minimal four-site model to distinguish it from the four-site model with an extra $SU(2)$ group studied elsewhere  \cite{Chivukula:2005bn,Accomando:2008dm,Accomando:2008jh,SekharChivukula:2008gz}.   In addition, as we elaborate in section \ref{sec:continuum} below, 
this $SU(2)^2 \times U(1)^2$ model
has a spectrum reminiscent of the most popular continuum Higgsless model.

In this section we analyze the $SU(2)_0 \times SU(2)_1 \times U(1)_2 \times U(1)_3$ linear moose shown in Fig.~\ref{fig:2-2}, and use the deconstruction sum rules to assess its ability to serve as a more complete low-energy effective theory for appropriate continuum models.   The Lagrangian for the link fields in this linear moose is
\begin{equation}
{\cal L} = \dfrac{f^2}{4} {\rm Tr} ( D_\mu \Sigma_1) (D^\mu \Sigma_1)^\dagger
+ \dfrac{f^2}{4} {\rm Tr} ( D_\mu \Sigma_2) (D^\mu \Sigma_2)^\dagger
+ \dfrac{f'^2}{8} {\rm Tr} (D_\mu \sigma) (D^\mu \sigma)^\dagger
\end{equation}
where the covariant derivatives of the link fields are given by
\begin{eqnarray}
D_\mu \Sigma_1 & = & \partial_\mu \Sigma_1 + i g W^a_{0\mu} \dfrac{\tau^a}{2} \Sigma_1 
- i \tilde{g} \Sigma_1 W^a_{1\mu} \dfrac{\tau^a}{2} \cr
D_\mu \Sigma_2 & = & \partial_\mu \Sigma_2 + i \tilde{g} W^a_{1\mu} \dfrac{\tau^a}{2} \Sigma_2 
- i g'  \Sigma_2 B_{2\mu} \dfrac{\tau^3}{2} \cr
D_\mu \sigma & = & \partial_\mu \sigma + i g' q B_{2\mu} \sigma - i \tilde{g}' \tilde{q} \sigma B_{3\mu}~,
\end{eqnarray}
where the $\sigma$ link field has charges $(q,\tilde{q})$ under $U(1)_2 \times U(1)_3$.
We work in the limit
\begin{equation}
 {\tilde g}, {\tilde g}' \gg g, g'~,
\end{equation}
so that the masses of the $W$ and $Z$ are much less than those of the heavy vector resonances
and, for simplicity, we consider the case that the decay constants of the two $\Sigma$ fields take 
the same value:
\begin{equation}
 f_1=f_2=f \equiv \sqrt{2} v~.
\end{equation}
The decay constant of $\sigma$, denoted by $f'$, is allowed to vary.  We introduce a leading-order weak mixing angle $\theta$, and dimensionless parameters 
$s, c, t,$ and $x$ as in the three-site model,
\begin{equation}
 \frac{g'}{g} = \frac{\sin{\theta}}{\cos{\theta}} 
\equiv \frac{s}{c} \equiv t\,, \qquad\qquad   \frac{g}{\tilde g} \equiv x \ \ (\ll 1) \,,
\end{equation}
and also a new parameter, $u$, describing the strength of the $U(1)_3$ coupling,
\begin{equation}
\frac{{\tilde g}'}{\tilde g} \equiv u.
\label{eq:txu}
\end{equation}

In the following subsections, we present the most salient properties of the gauge and fermion sectors, stressing comparisons with the three-site model; further detail is given in Appendices A and B.  We then look in greater depth at the $Z'$ and $Z''$ bosons in comparison to the heavy neutral gauge bosons of the three-site model and continuum models.

\subsection{The gauge sector}

Because the $SU(2) \times SU(2)$ gauge sector of this model is identical to that of the three-site model \cite{SekharChivukula:2006cg}, the charged-gauge boson mass-squared matrix is unaltered.   The $W$ and $W'$ boson masses and wavefunctions and the expression for $G_F$ in terms of the model parameters ($g$, $v$, $x$) are the same as in Ref. \cite{SekharChivukula:2006cg}.  

The mass-squared matrix for the neutral gauge bosons, on the other hand, is enlarged by the addition of the second $U(1)$ group. The matrix has a zero eigenvalue, corresponding to the massless photon, with corresponding electric charge 
\begin{equation}
 \frac{1}{e^2} = \frac{1}{g^2} + \frac{1}{{\tilde g}^2}
             + \frac{1}{g'^2} + \frac{q^2}{ \tilde{q}^2 \tilde{g}'^2}~.
\label{eq:e}
\end{equation}
As we show in the next subsection, we will require $q=\tilde{q}=1/6$ in order to provide
appropriate hypercharges to the light fermions. Therefore, in what follows, we will assume
$q=\tilde{q}$, in which case the factors of $q$ and $\tilde{q}$ in the last term of Eq. (\ref{eq:e}) cancel, yielding:
\begin{equation}
\frac{1}{e^2} = \frac{1}{g^2 s^2}\left[1+s^2\left(1+\frac{1}{u^2}\right)x^2 \right].
\end{equation}
The light neutral gauge boson, which we associate with the 
$Z$, has a mass
\begin{equation}
 M_Z^2 = \frac{g^2 v^2}{4}\left[(1+t^2)  
 - \left(\frac{(1-t^2)^2}{4}+\frac{t^4}{u^2} \right)x^2 + O(x^4)\right] \,.
\label{eq:Zmass}
\end{equation}
In the limit as $u\to \infty$, the expressions describing the photon and the $Z$ boson recover the values of the three-site model.   The neutral gauge sector also includes two heavy mass eigenstates instead of the single $Z'$ boson of the three-site model; we will discuss these states further in  \ref{subsec:zprimme}

\subsection{$Z$-standard weak mixing angle and multi-gauge-boson couplings}

To facilitate comparison of this model with experiment, we take $M_Z$ as an input and define a ``$Z$-standard" weak mixing angle $\sin\theta_Z \equiv s_Z$ in terms of $e$, $M_Z$ and $G_F$:
\begin{eqnarray}
 s_Z^2 c_Z^2 &\equiv& \frac{e^2}{4\sqrt{2} G_F M_Z^2} \nonumber \\
 &=& s^2c^2\left[ 1 + (c^2-s^2)\left(\left(1-\frac{1}{4c^2}\right) 
 - \frac{t^2}{u^2} \right)
 x^2 
 + O(x^4) \right] \,.
 \label{eq:zstdmx}
\end{eqnarray}
The relation between the weak mixing angle $\sin\theta_Z$ and the quantity $s$ defined in Eq.~(\ref{eq:txu}) is expressed as follows:
\begin{equation}
 s_Z^2 = s^2 + \Delta,\ \ \ c_Z^2 = c^2 - \Delta,
\label{eq:Zstandard}
\end{equation}
\begin{equation}
 \Delta = \left[ s^2 \left(c^2-\frac{1}{4}\right) 
 -\frac{s^4}{u^2} \right] x^2 + O(x^4).
\label{eq:Delta}
\end{equation}

Several quantities that depend on the new $U(1)$ coupling through the parameter $u$ when written in terms of $s$ are $u$-independent to leading order when written in terms of $s_Z$.   For example, the coupling $g$ in Eq. (\ref{eq:e}) has an explicit dependence on $u$ when written in terms of $s$
\begin{equation}
 g^2 = \frac{e^2}{s^2}\left[1+s^2\left(1+\frac{1}{u^2}\right)x^2\right] = \frac{e^2}{s_Z^2} \left[1+\frac{3}{4}x^2 + O(x^4)
 \right] \,, \label{eq:gsqu}
\end{equation}
but recovers the three-site form \cite{SekharChivukula:2006cg} when expressed in terms of $s_Z$.
Similarly, the $u$ dependence of $v_Z^0$ and $v_Z^1$ visible in eqs. (\ref{eq:vZ0}) and (\ref{eq:vZ1}) disappears when we re-express these quantities in terms of $s_Z$:
\begin{eqnarray}
v_Z^0 &=& c_Z \left[ 1 - c_Z^2 \frac{1-t_Z^2-2t_Z^4}{8}x^2 + O(x^4) \right]~, \\
v_Z^1 &=& \frac{c_Z x}{2} \left(1 - \frac{s_Z^2}{c_Z^2}\right) - \frac{c_Z^3 x^3}{16} \left(1 - \frac{s_Z^2}{c_Z^2}\right)^3 + O(x^5)~.
\end{eqnarray}
These are again the same form as in the three-site model \cite{SekharChivukula:2006cg}.  As a result, the $ZWW$, $WWWW$, and $ZW'W$ gauge couplings take on their three-site model values when written in terms of $s_Z$:
\begin{eqnarray}
 g_{ZWW} &=& g (v_W^0)^2 (v_Z^0) + {\tilde g} (v_W^1)^2 (v_Z^1) = e\frac{c_Z}{s_Z}\left[1 + \frac{1}{8 c_Z^2}x^2 + O(x^4)\right],\label{eq:ZWW} \\
 g_{ZW'W} &=& g (v_W^0)(v_{W'}^0) (v_Z^0) + {\tilde g} (v_W^1) (v_{W'}^1) (v_Z^1) = - \frac{ex}{4 s_Z c_Z} \left[1 + \frac{s_Z^2}{4 c_Z^2}x^2 + O(x^4) \right],\label{eq:ZWWp}\\
 g_{WWWW} &=& g^2 (v_W^0)^4 + {\tilde g}^2 (v_W^1)^4
 = \frac{e^2}{s_Z^2}\left[1 + \frac{5}{16}x^2 + O(x^4)\right]. \label{eq:WWWW}
\end{eqnarray}
For $g_{ZWW}$ and $g_{ZW'W}$, this is because the terms to $O(x^3)$ are independent of $u$ and for $g_{WWWW}$ it is because the $W$ boson itself does not differ from that of the three-site model. 

The values of $g_{ZWW}$ and $g_{WWWW}$ are sufficient to determine the chiral Lagrangian \cite{Appelquist:1980ix,Gasser:1983yg,Appelquist:1993ka} coefficients $\alpha_i, \ [i = 1...5]$ and to ensure that these, too, have their three-site model values.  The additional four-gauge-boson couplings $g_{WWZZ}$ and $g_{WWZ\gamma}$ can be determined in terms of the $\alpha_i$ and shown to have the values:
\begin{eqnarray}
g_{WWZZ} &=& g^2 (v_W^0)^2(v_Z^0)^2  + \tilde{g}^2 (v_W^1)^2(v_Z^1)^2
= e^2\frac{c_Z^2}{s_Z^2}\left[1 + \left( \frac{1}{4 c_Z^2} + \frac{1}{16 c_Z^4} \right) x^2 + O(x^4)\right],
\label{eq:gWWZZ} \\
g_{WWZ\gamma} &=& g^2 (v_W^0)^2(v_Z^0)(v_\gamma^0) 
          + \tilde{g}^2 (v_W^1)^2(v_Z^1)(v_\gamma^1)
          = e^2\frac{c_Z}{s_Z}\left[1 +  \frac{1}{8 c_Z^2}x^2 + O(x^4)
    \right] \,,
\label{eq:gWWZA}
\end{eqnarray}
which again have no $u$-dependence and also apply in the three-site model (they were not derived in Ref. \cite{SekharChivukula:2006cg}).

In contrast, the properties of the $Z'$ and $Z''$ bosons distinctly differ from those of the three-site model $Z'$ boson.  We will return to this topic after discussing the fermion sector.

\subsection{The fermion sector}

\begin{figure}[hbt]
  \begin{center}
    \includegraphics[height=4cm]{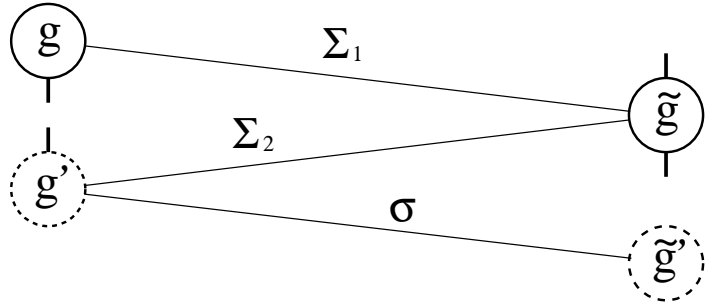}
  \end{center}
\caption{A folded moose diagram representing the same model as in Fig.~\ref{fig:2-2}.  As noted in the text, the hypercharge fermionic currents at site 0 and site 2 couple to the $U(1)$ gauge boson at site 2, while the hypercharge fermionic currents at site 1 couple to the $U(1)$ gauge boson at site 3. 
}
\label{fig:folded}
\end{figure}

The fermion sector has delocalized $U(1)$ charges as well as delocalized $SU(2)$ charges. As shown in Table II, 
\begin{table}[bt]
\begin{center}
\begin{tabular}{|c||c|c|c|c|c|c|}
\hline\hline &&&&&&\\
& \ $Q_{L0}$,\ & $Q_{L1}$, $Q_{R1}$, & $u_{R2}$, $d_{R2}$, & \ $\Sigma_1$\  & \ $\Sigma_2$\  & \  $\sigma$\  \ \ \\ [2mm]
& $L_{L0}$  &  $L_{L1}$, $L_{R1}$ & $\nu_{R2}$, $\ell_{R2}$ & & & \\
\hline\hline &&&&&&\\
$SU(2)_0$& \bf{2}& \bf{1}& \bf{1}&  \bf{2} & \bf{1} & \bf{1}\\[2mm]
\hline &&&&&&\\
$SU(2)_1$& \bf{1}&  \bf{2}& \bf{1}& \bf{2} & \bf{2} & \bf{1} \\ [2mm]
\hline &&&&&&\\
$U(1)_2$ & $Y_{SM}$ &0 & $Y_{SM}$ & 0 & $\begin{pmatrix} +\frac12&  \cr  & -\frac12 \end{pmatrix}$& $\dfrac16$ \\ [3mm]                   
\hline &&&&&&\\
$U(1)_3$ &  0 & $\frac16$ (Q); $-\frac12$ (L) & 0 & 0 & 0 &  $\dfrac16$\\ [2mm]
\hline\hline
\end{tabular}
\caption{Gauge charge assignments of the fermions and Nambu-Goldstone bosons under the weak and hypercharge groups of the minimal four-site model. The notation $Y_{SM}$ refers to the hypercharge value of the related state in the Standard Model; we use the convention $Q = T_3 + Y$ so that the hypercharge of a left-handed quark doublet is $\frac{1}{6}$ and that of a left-handed lepton doublet is $-\frac{1}{2}$.}
\end{center}
\label{tab:chargeswredux}
\end{table}
we take the fermions at site 0 ($\psi_{L0}$) and site 2 ($u_{R2}$, $d_{R2}$) to be charged under $U(1)_2$, while the fermions at site 1 ($\psi_{L1}$, $\psi_{R1}$) are charged under $U(1)_3$.
This choice of fermion charges allows us to interpret the model as a highly-deconstructed version of a 5d continuum $SU(2) \times SU(2) \times U(1)$ gauge theory, as illustrated by the folded moose diagram in Fig.~\ref{fig:folded}.  The light fermions and the gauge groups under which they are charged lie together on the``UV brane" side of the Moose, while the vector-like fermions and their gauge groups are together on the``TeV brane" side.  Table II also lists the gauge charges of the Nambu-Goldstone modes; the $U(1)$ charges of these states were chosen to enable the light fermion mass eigenstates to have the expected standard-model-like values of hypercharge.

The Lagrangian terms that contribute to the fermion masses (Yukawa coupling terms and Dirac mass terms) resemble those in the three-site model -- with the key difference that the Yukawa terms now include the Nambu-Goldstone mode, $\sigma$, of the third link.   In 5d language, the Yukawa terms arise from the covariant derivative along the 5th dimension; hence they include the pions of both the $SU(2)$ and $U(1)$ gauge fields
\begin{eqnarray}
{\cal L}_{quark} &=& M \left[ \varepsilon_{L} \bar{Q}_{L0} \Sigma_1 \sigma Q_{R1} + \bar{Q}_{R1} Q_{L1} + 
\bar{Q}_{L1} \Sigma_2
\begin{pmatrix}
\varepsilon_{uR} & \\
& \varepsilon_{dR}
\end{pmatrix} \sigma^*
\begin{pmatrix}
u_{R2} \\
d_{R2}
\end{pmatrix}
+ \textrm{h.c.}\right]~, \\
{\cal L}_{lepton}&=& M \left[ \varepsilon_{L} \bar{L}_{L0} \Sigma_1 (\sigma^*)^3 L_{R1} + \bar{L}_{R1} L_{L1} + 
\bar{L}_{L1} \Sigma_2 
\begin{pmatrix}
\varepsilon_{\nu R} & \\
& \varepsilon_{\ell R}
\end{pmatrix}
(\sigma)^3
\begin{pmatrix}
\nu_{R2} \\
\ell_{R2}
\end{pmatrix}
+ \textrm{h.c.}\right]~.
\label{eq:yukawa}
\end{eqnarray}
Note that while the additional sigma field, $\sigma$, is part of the quark Yukawa terms, it is  the cube of $\sigma^*$ that is part of the lepton Yukawa terms, since the hypercharges of the left-handed quarks and leptons differ by a factor of negative three.\footnote{More generally, one could allow the charges of $\sigma$ under the two $U(1)$ groups to be different and could allow different powers of $\sigma$ or $\sigma^*$ to appear in the various Lagrangian terms:
\begin{eqnarray*}
{\cal L}_{quark} &=& M \left[ \varepsilon_{L} \bar{Q}_{L0} \Sigma_1 (\sigma)^r Q_{R1} + \bar{Q}_{R1} Q_{L1} + 
\bar{Q}_{L1} \Sigma_2
\begin{pmatrix}
\varepsilon_{uR} & \\
& \varepsilon_{dR}
\end{pmatrix} (\sigma^*)^{r'}
\begin{pmatrix}
u_{R2} \\
d_{R2}
\end{pmatrix}
+ \textrm{h.c.}\right] \\
{\cal L}_{lepton}&=& M \left[ \varepsilon_{L} \bar{L}_{L0} \Sigma_1 (\sigma^*)^{s'} L_{R1} + \bar{L}_{R1} L_{L1} + 
\bar{L}_{L1} \Sigma_2 
\begin{pmatrix}
\varepsilon_{\nu R} & \\
& \varepsilon_{\ell R}
\end{pmatrix}
(\sigma)^s
\begin{pmatrix}
\nu_{R2} \\
\ell_{R2}
\end{pmatrix}
+ \textrm{h.c.}\right] 
\label{eq:yukawa-general}
\end{eqnarray*}
Requiring the light fermions to have the same hypercharges as their SM counterparts turns out to impose the constraints $q = \tilde{q}$, $3r = 3r' = s = s'$, and $qr = \frac16$.   Our choice in the text of $q = \frac16$ and $r = 1$ represents the largest  value of $q$ for which $r$ is an integer.}

The mass matrices for the quarks and leptons are the same as in the three-site model.  Hence the
expressions for the light left-handed and right-handed fermion mass eigenstates are, likewise, as in the three-site model, when written in terms of the Yukawa-like couplings $\varepsilon_L$ and $\varepsilon_R$.  
Since the couplings of the light fermion eigenstates to the $W$ arise from the overlap of the fermion and $W$ eigenstates, these also remain as in the three-site model.  For example, the right-handed coupling of the third-generation quarks to the $W$ is
\begin{equation}
g^{Wtb}_R = \tilde{g}\,t^1_R\,b^1_R\,v^1_W = \frac{g}{2} \frac{\varepsilon_{tR}}{\sqrt{1+\varepsilon^2_{tR}}} \frac{\varepsilon_{bR}}{\sqrt{1+\varepsilon^2_{bR}}} \left[1+O(x^2)\right] \\
\approx \frac{g}{2}\frac{m_b}{m_t} \frac{\varepsilon_{tR}}{1+\varepsilon^2_{tR}}\,,
\label{eq:gWRtb}
\end{equation}
and because this has the same form as in the three-site model, the value of the upper bound on 
$\varepsilon_{tR}$ as constrained by $b \rightarrow s \gamma$ 
is unchanged, i.e., $\varepsilon_{tR}<0.67$.

 In any linear Higgsless model, there exists a  particular value of $\varepsilon_L$ such that the left-handed fermion profile ($f^i_L$) is related to that of the $W$ boson ($v^i_W$) as
\begin{equation}
g_i (f^i_L)^2 \propto v^i_W\,,
\end{equation}
and is said to exhibit ``ideal delocalization'' \cite{SekharChivukula:2005xm}.  As a result, since the $W$ and $W'$ eigenstates are orthogonal, the fermions are rendered incapable of coupling to  the $W'$, thereby eliminating the possibility of large precision electroweak corrections\footnote{In principle, there is a correction to the parameter Y \cite{Barbieri:2004qk} because of the additional $U(1)$ gauge boson. However, in the presence of ideal delocalization, the correction to $Y$ begins at order $x^4$ \cite{SekharChivukula:2005xm} and is therefore negligible.} from fermion loops, which would conflict with existing data.  Since both the $W$ boson profile and the expressions for the fermion eigenstates are as in the three-site model, the value of $\varepsilon_L$ corresponding to ideal delocalization is unchanged in our four-site model:
\begin{equation}
\varepsilon_L = \frac{x^2}{2} + {\cal{O}}(x^4) \,.
\end{equation}

Most other properties of the ideally-delocalized fermion sector in this four-site model are nearly identical to those of the three-site model.  For example, both the weak and hypercharge couplings of light fermions to the $Z$ boson are unchanged and are, in fact, essentially the same as in the Standard Model:
\begin{equation}
 g_{3L}^{Zqq} =  \frac{eM_W}{M_Z\sqrt{1-\frac{M_W^2}{M_Z^2}}}
 \left[1+O(x^4)\right]\,, \qquad\qquad\qquad
 g_{YL}^{Zqq} = -\frac{eM_Z}{M_W}\sqrt{1-\frac{M_W^2}{M_Z^2}}
 \left[1+O(x^4)\right]\ .
 \label{eq:foursiteferms}
\end{equation}
Details on the couplings of the fermions to the $Z$ are given in Appendix A.  In particular, the hypercharge coupling of the right-handed top quark to the $Z$ is slightly altered because there are contributions from both site 2 and site 3 -- and the $Z$ boson profile is not the same at both sites.

\subsection{The $Z'$ and $Z''$ bosons}
\label{subsec:zprimme}

We now return to considering the properties of the heavy neutral gauge bosons $Z'$ and $Z''$ in the four-site model.   The mass-squared matrix for the neutral gauge bosons (displayed here for arbitrary $\sigma$ charges) is enlarged by the addition of the second $U(1)$ group and takes the form
\begin{equation}
\frac{{\tilde g}^2 v^2}{2}
\left(
\begin{array}{cccc}
\ \ x^2\ \ &\ \ -x\ \ &0&0\\
\ \ -x\ \ &\ \ 2\ \ &\ \ -xt\ \ &0\\
0&\ \ -xt\ \ &x^2t^2 \left(1 + q^2 \left[\dfrac{f'}{f}\right]^2\right)&-xtu\,  q \tilde{q} \left[\dfrac{f'}{f}\right]^2\\
0&0&-xtu\, q \tilde{q} \left[\dfrac{f'}{f}\right]^2&u^2\, \tilde{q}^2 \left[\dfrac{f'}{f}\right]^2 
\end{array}
\right).
\label{eq:mass_matrix-prime}
\end{equation}
A complete perturbative analysis of the eigenstates and eigenvalues of this system is provided
in Appendix A, but it is important to note that when
\begin{equation}
u^2\, \tilde{q}^2 \left[\dfrac{f'}{f}\right]^2 =2
\label{eq:degenerate}
\end{equation}
the heavy $Z'$ and $Z''$ bosons are approximately degenerate in the small-$x$ 
limit. The form of the masses and wavefunctions are different in the degenerate case
than in general -- and we therefore consider these cases separately below.

\subsubsection{The degenerate case: $u^2\tilde{q}^2 \left[\dfrac{f'}{f}\right]^2= 2$}

First, we consider the approximately degenerate case.  For the reasons discussed in
the previous section we take $q=\tilde{q}$, but for clarity we display the dependence on the
value of $\tilde{q}$ explicitly; in addition, we use the ``degeneracy condition" (Eqn. (\ref{eq:degenerate})) to eliminate $f'/f$.
Finishing the diagonalization of the mass-squared matrix (\ref{eq:mass_matrix-prime}) yields the following masses of the $Z'$ and $Z''$ bosons (the wavefunctions are given in Appendix A):
\begin{equation}
 M^2_{Z' (Z'')} = \tilde{g}^2 v^2 
   \left[1 + \frac{1}{8u^2} \left( u^2 + t^2(4+u^2) \mp w \right)x^2 + O(x^4)  \right], 
   \label{eq:heavyZmass_u-sqrt2} 
 \end{equation}
where
\begin{equation}
w \equiv \sqrt{u^4+2t^2u^2(-4+u^2)+t^4(4+u^2)^2}~.
\end{equation}
In this case, the two heavy bosons are nearly degenerate and their wavefunctions at site 1 and site 3 are both of $O(1)$ while those at site 0 and site 2 are of $O(x)$.

We may also compute the coupling of each heavy neutral gauge boson to a pair of $W$ bosons, using 
Eqs.~(\ref{eq:Wwave}) and (\ref{eq:heavyZwave_usq2}):
\begin{eqnarray}
 g_{Z'WW (Z''WW)} &=& g (v_W^0)^2 v_{Z' (Z'')}^0 + \tilde{g} (v_W^1)^2 v_{Z' (Z'')}^1 \label{eq:teripple} \\
  &=& \frac{e}{2 s_Z}\frac{\sqrt{2}\, t^2\, u}{\sqrt{w\{w\pm(u^2+t^2[-4+u^2])\}}} x +O(x^3).
\end{eqnarray}
While each of these couplings is significantly different at leading order than the $Z'WW$ coupling in the three-site model, there is a relation between the three couplings:
\begin{equation}
 \left(  g_{Z'WW}^2 + g_{Z''WW}^2\right)_{\rm four-site} 
 = \left( g_{Z'WW}^2 \right)_{\rm three-site} = \frac{e^2 x^2}{16 s_Z^2}\left[ 1+O(x^2) \right]  .
\label{eq:rel_3_4}
\end{equation}
In fact, this relationship is independent of the value of $u$ and follows directly from the general sum rule  (\ref{eq:E4sum})  that is responsible for ensuring the vanishing of the $E^4$ growth of the amplitude for $W_L W_L \to W_L W_L$ scattering because $g_{\gamma WW}$ is fixed by gauge invariance and we showed earlier (Eqs. (\ref{eq:ZWW}) and (\ref{eq:WWWW})) that $g_{WWWW}$ and $g_{ZWW}$ are identical in the three-site
 and 4-site models.
 
Given the relationship among the couplings (\ref{eq:rel_3_4}), the near-degeneracy of the $Z'$ and $Z''$ boson masses in this case, and the near-degeneracy of those masses with the $Z'$ boson mass in the three-site model, we find that the expression (\ref{eq:thirdahat}) for $\hat{a}$ reduces to
\begin{eqnarray}
\hat{a}^{4-site}_{degenerate}& =& \frac{v^2}{M_W^4}
 (g_{Z'WW}^2 M_{Z'}^2 + g_{Z''WW}^2M_{Z''}^2 )_{\rm four-site, degenerate} \nonumber \\
 & =& \frac{v^2}{M_W^4} \left( [g_{Z'WW}^2  + g_{Z''WW}^2] M_{Z'}^2 \right)_{\rm four-site, degenerate} \\
 &\approx&  \frac{v^2}{M_W^4} ( g_{Z'WW}^2  M_{Z'}^2)_{\rm three-site}  = 1\,, \nonumber
\end{eqnarray}
so that  the ``degenerate" four-site model still has $\hat{a} \approx 1$, just like the three-site model.

\subsubsection{The non-degenerate case: $u^2\tilde{q}^2 \left[\dfrac{f'}{f}\right]^2 - 2 = {{O}(1)}$ }

We now consider the alternative limit, without degeneracy.   Finishing the diagonalization of the mass-squared matrix (\ref{eq:mass_matrix}) yields the following masses for the $Z'$  and $Z''$ if one makes the assumption that $(u^2\, \tilde{q}^2 \left[f'/f\right]^2-2)={{O}(1)}$ (again, the wavefunctions are listed in Appendix A, and we take $q=\tilde{q}$):
\begin{eqnarray}
 M_{Z'}^2 &=& \tilde{g}^2 v^2\left[1  
 + \frac{1}{4}\left(1+t^2\right)x^2 + O(x^4)\right],
 \label{eq:Z'mass}\\
  M_{Z''}^2 &=& \frac{\tilde{g}^2 v^2}{2}\left[\frac{f'}{f}\right]^2\,\tilde{q}^2\left[u^2 
 + t^2x^2 + O(x^4)\right] \,.
\label{eq:Z''mass}
\end{eqnarray}
This time, the ratio of masses is $\left(M_{Z''}/M_{Z'}\right)^2 = \frac{u^2 \tilde{q}^2}{2}\left[\frac{f'}{f}\right]^2[1+O(x^2)] $; we also find
that the $Z'$ boson is strongly concentrated at site 1 and the $Z''$ boson, at site 3.  In the $u \rightarrow \infty$ limit, we recover the three-site model: the $Z'$ mass and wavefunction revert to the three-site form, while the $Z''$ becomes infinitely massive and localized at site 3.  

We may also compute the coupling of the $Z'$ and $Z''$ to a pair of $W$ bosons, as in Eq. (\ref{eq:teripple}).  The results are:
\begin{eqnarray}
g_{Z'WW} &=&  - \frac{e\,x}{4 s_Z} + O(x^3) \,,  \label{eq:4zpww} \\
g_{Z''WW} &=&  \frac{e s_Z x^3}{4 c_Z^2}
\left(\frac{f^2}{{f'}^2\tilde{q}^2u^3}\right)
\left(\frac{4 f^2-{f'}^2\tilde{q}^2u^2}{2f^2-{f'}^2 \tilde{q}^2 u^2}\right)
 + O(x^5) \, , \label{eq:4zppww}
\end{eqnarray}
where, as before, we take $q=\tilde{q}$ and explicitly display the $\tilde{q}$ dependence.
As expected, in the limit where $u \to \infty$, the $Z'WW$ coupling takes on its three-site value and $Z''WW$ coupling duly vanishes; this is consistent with having the $Z'$ and $Z''$ bosons increasingly localized on sites 1 and 3, respectively.  In addition, since $g_{Z''WW} =
{\cal O}(x^3)$, the sum of the squares of these couplings behaves as in Eq. (\ref{eq:rel_3_4}), as mandated by the general sum rule (\ref{eq:E4sum}).  

Finally, evaluating Eq. (\ref{eq:thirdahat}) by applying Eqs. (\ref{eq:M_W2}), (\ref{eq:gsqu}), (\ref{eq:Z'mass}), (\ref{eq:Z''mass}),  (\ref{eq:4zpww}), and (\ref{eq:4zppww}), we find that $\hat{a} = 1$ when $u^2\tilde{q}^2 [{f'/f}]^2 \neq 2$.  The contribution from the $g^2_{Z'} M_{Z'WW}$ dominates and that from $g^2_{Z''} M_{Z''WW}$ is suppressed by a factor of $x^4$. In other words, the parameter $\hat{a}$ equals 1 in the minimal four-site model, independent of the value of $u$.

\subsection{Comparison to a continuum  $SU(2) \times SU(2) \times U(1)$ 5d Higgsless model}
\label{sec:continuum}

By comparison, let us consider the form of the neutral heavy gauge bosons  in a continuum 
$SU(2)^2 \times U(1)$ Higgsless model with a flat extra dimension.  We will call the coordinate of the extra dimension $y$ and work in the unfolded picture in which $y$ runs from $0$ to $3\pi R$.  Using the methods discussed in Ref.~\cite{Chivukula:2005cc}, the expressions for masses and wavefunctions for heavy gauge bosons that result from  solving the mode equations with boundary conditions are,
\begin{eqnarray}
 \tilde{M}_{Z', Z''} &\equiv&  \pi R M_{Z', Z''}   \nonumber \\
 &=&
  \frac{\pi}{2} 
  + \left( \frac{2t_0^2+r^2(1+t_0^2)\pm\sqrt{(2t_0^2+r^2(1+t_0^2))^2-8r^2t_0^2}}{\pi r^2} \right) \lambda 
  + O(\lambda^2) ,
\label{eq:heavyZmass_cont}
\end{eqnarray}
\begin{eqnarray}
  \chi_{Z', Z''}(\tilde{y}) &=& 
  C_{Z', Z''} 
  \left[  
   \cos(\tilde{y}\tilde{M}_{Z', Z''})  
   - \frac{\tilde{M}_{Z', Z''}}{2\lambda}\sin(\tilde{y}\tilde{M}_{Z', Z''})
  \right] \hspace{3.4cm} \mbox{(for $0 \le \tilde{y} \le 2$)}\, , \\
  &=& 
  C_{Z', Z''} 
  \left[  
  \frac{\cos(2\tilde{M}_{Z', Z''}) - \frac{\tilde{M}_{Z', Z''}}{2\lambda}\sin(2\tilde{M}_{Z', Z''})}
      {\cos(\tilde{M}_{Z', Z''})}
  \right] \,
   \cos((3-\tilde{y})\tilde{M}_{Z', Z''}) \ \ \ \ \mbox{(for $2 \le \tilde{y} \le 3$)}\, ,
\label{eq:heavyZwave_u-sqrt2}
\end{eqnarray}
where, $C_{Z', Z''}$ are normalization constants for the $Z'$ and $Z''$ wavefunctions,
$t_0$ and $\lambda$ are defined in Eqs. (\ref{eq:t0}) and (\ref{eq:lambda}), $r=g_{5Y}/g_{5}$
is the ratio of the five-dimensional hypercharge and $SU(2)$ couplings,
and $\tilde{y} \equiv y/(\pi R)$.  From Eq. (\ref{eq:heavyZmass_cont}) we see that the lowest-lying pair of KK modes, $Z'$ and $Z''$ are almost degenerate, i.e., 
${M_{Z'}}/{M_{Z''}} = 1 + O(\lambda)$.  Moreover,  the wavefunction of each boson has similar magnitude for $\tilde{y} = 1,3$ and a much smaller magnitude for $\tilde{y} = 0,2$.   This suggests that the deconstructed four-site model will best approximate the behavior of the heavy gauge bosons in the $SU(2)\times SU(2) \times U(1)$ continuum model for the parameter value 
$u^2\tilde{q}^2 \left[f'/f\right]^2\approx 2$.  

Recall that the charged electroweak bosons of the four-site and three-site models are identical, while the pair of nearly-degenerate $Z'$ and $Z''$ bosons for $u^2\tilde{q}^2 \left[f'/f\right]^2=2$ jointly couple to $W$-boson pairs like the single $Z'$ boson of the three-site model.  We therefore conclude that vector-boson scattering processes among the lightest KK gauge bosons in a continuum $SU(2)^2 \times U(1)$ model with a flat extra dimension will be well-described by the three-site model.  Since the three-site model also provides an effective low-energy description of the $SU(2)\times SU(2)$ continuum model, it follows that the lightest KK gauge modes of the  $SU(2)^2 \times U(1)$ and $SU(2)^2$ continuum models will also appear phenomenologically similar at low energies.

\section{The Triangular Moose Model with Three Sites and Three Links}

\begin{figure}
\begin{center}
\includegraphics[scale=1.0]{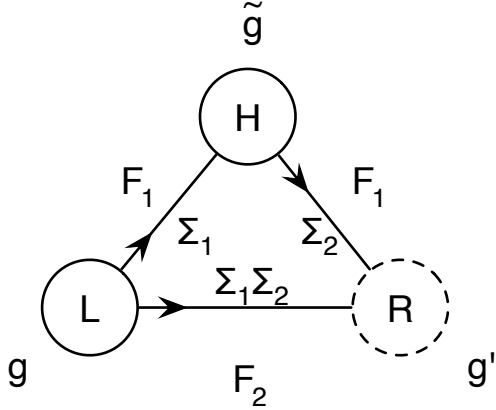}
\end{center}
\caption{\label{fig:triangle} A moose diagram of the $SU(2)^2 \times U(1)$ hidden local symmetry (HLS) \protect\cite{Casalbuoni:1985kq,Casalbuoni:1996qt,Bando:1985ej,Bando:1985rf,Bando:1988ym,Bando:1988br,Harada:2003jx} Higgsless model 
discussed in this section. As discussed in the text, the model includes only two separate $SU(2)\times SU(2)/SU(2)$ nonlinear sigma model fields, $\Sigma_{1,2}$. We will work in the limit $g,g' \ll \tilde{g}$,
where $g$ and $g'$ are the coupling constants of the $SU(2) \times U(1)$ gauge
groups labeled $L$ and $R$ respectively (and the $U(1)$ is generated
by $T_3$ in the global $SU(2)_R$), and $\tilde{g}$ is the coupling of the $SU(2)_H$ group
at the apex of the diagram. The fermions derive their couplings
from {\it both} $SU(2)$ groups, as required by ideal delocalization 
\protect\cite{SekharChivukula:2005xm}.}
\end{figure}

Having analyzed the minimal four-site model, we now consider a different modifications of the three-site model \cite{SekharChivukula:2006cg}  that can remedy some of its shortcomings, as measured by the deconstruction sum rules.  Because Ref. \cite{SekharChivukula:2008mj} discussed how to extend the relevant sum rules to ring models, the second extension we consider is a general breaking electroweak symmetry strongly (BESS) or hidden local symmetry (HLS) model \cite{Bando:1985ej,Bando:1985rf,Bando:1988ym,Bando:1988br,Harada:2003jx}, a three-site moose model that includes an additional link closing the moose into a ring (see Fig. \ref{fig:triangle}).  In order to have a consistent electroweak
phenomenology, we incorporate ideal fermion delocalization in section \ref{sec:fermiontriangle} 
below and we refer to the complete model as the triangular moose from here on.

In this section we describe the triangular moose model and compare its properties to those of the three-site model and appropriate continuum theories.  This model\footnote{The heavy fermion phenomenology of a model with a similar gauge structure but a different top quark sector is studied in \cite{Chivukula:2009ck}.}, as illustrated in  Fig. \ref{fig:triangle}, is based on an $SU(2)_L \times SU(2)_H \times U(1)_R$ gauge group, with the $SU(2)_L \times U(1)_R$ couplings $g$ and $g'$ much less than
the $SU(2)_H$ coupling $\tilde{g}$. The $SU(2)_H$ represents the deconstructed
bulk gauge fields of a five-dimensional Higgsless model and the gauged $U(1)_R$ corresponds to the subgroup generated by $T_3$ in a  global $SU(2)_R$.  The model incorporates
two nonlinear sigma model fields $\Sigma_{1,2}$ which transform as 
\begin{equation}
\Sigma_1 \to L \Sigma_1 H^\dagger~, \hskip+2.0cm
\Sigma_2 \to H \Sigma_2 R^\dagger~,
\end{equation}
under the global $SU(2)_L \times SU(2)_H \times SU(2)_R$ symmetry, so their covariant derivatives are:
\begin{eqnarray}
\label{eq:triagain}
D^\mu \Sigma_1 &= & \partial^\mu \Sigma_1 + ig W_L^{a\mu} \frac{\tau^a}{2}\,\Sigma_1 - i \tilde{g} \Sigma_1 W_H^{a\mu} \frac{\tau^a}{2}~,  \\
D^\mu \Sigma_2 &= & \partial^\mu \Sigma_2 +  i \tilde{g} W_H^{a\mu} \frac{\tau^a}{2}  \Sigma_2 - i g' \Sigma_2 W_{3R}^\mu \frac{\tau^3}{2}~.
\end{eqnarray}

At ${\cal O}(p^2)$, the most general parity-invariant
Lagrangian for $\Sigma_{1,2}$ is given 
by
\begin{equation}
{\cal L}^{triangular}_{p^2} = \frac{F_1^2  }{4} {\rm Tr} \left[ (D_\mu \Sigma_1)^\dagger(D^\mu \Sigma_1)  +   (D_\mu \Sigma_2)^\dagger(D^\mu \Sigma_2) \right]  + \frac{F_2^2}{4} \,  {\rm Tr} \left[ (D_\mu (\Sigma_1\Sigma_2))^\dagger  D^\mu (\Sigma_1\Sigma_2)   \right]\,.
\label{eq:triangle-Lag}
\end{equation}
There are also kinetic energy
terms for the gauge-bosons
\begin{equation}
{\cal L}_{GB} = -\frac{1}{4}\left[
(\vec{W}^{\mu\nu}_L)^2 +(\vec{W}^{\mu\nu}_H)^2+(W^{\mu\nu}_{3R})^2  
\right]~.
\label{eq:gblag}
\end{equation}
The triangular moose gauge sector  is equivalent to the BESS or HLS Lagrangian given in refs. \cite{Bando:1985ej,Bando:1985rf,Bando:1988ym,Bando:1988br,Harada:2003jx}
\begin{align}
{\cal L}^{HLS} = & -\,\frac{v^2}{4} {\rm Tr} \left[(D^\mu \Sigma^\dagger_1) \Sigma_1
- (D^\mu \Sigma_2) \Sigma^\dagger_2\right]^2  -a\,\frac{v^2}{4} {\rm Tr}\,\left[(D^\mu \Sigma^\dagger_1) \Sigma_1 + (D^\mu \Sigma_2) \Sigma^\dagger_2\right]^2~ \nonumber \\
=& \frac{v^2}{4}(1+a) {\rm Tr} \left[ (D_\mu \Sigma_1)^\dagger(D^\mu \Sigma_1)  +   (D_\mu \Sigma_2)^\dagger(D^\mu \Sigma_2)  \right] + \frac{v^2}{2}(1-a) {\rm Tr} \left[ (D_\mu \Sigma_1)^\dagger \Sigma_1 (D_\mu \Sigma_2) \Sigma_2^\dagger  \right] \nonumber \\
=& \frac{a v^2 }{2} {\rm Tr} \left[ (D_\mu \Sigma_1)^\dagger(D^\mu \Sigma_1)  +   (D_\mu \Sigma_2)^\dagger(D^\mu \Sigma_2) \right]  + \frac{v^2}{4}(1-a) \,  {\rm Tr} \left[ (D_\mu (\Sigma_1\Sigma_2))^\dagger  D^\mu (\Sigma_1\Sigma_2)   \right]\,.
\label{eq:HLS}
\end{align}
if we establish the correspondence 
\begin{align}
F^2_1 & = 2 a v^2 ~,
\label{eq:tri-HLS-corresp}\\ \nonumber
F^2_2 & = v^2(1-a)~.
\end{align}
Note that, for $a \neq 1$, this Lagrangian is not ``local in theory space"
\cite{ArkaniHamed:2001ca}. As discussed below, the case where $a=\frac43$
is particularly interesting since it most closely approximates\footnote{Note 
that $a=4/3$ corresponds to $\rho$-meson dominance of $\pi\pi$ scattering in 
HLS models of QCD \protect\cite{Harada:2003jx,Harada:2006di}.} the properties
of a continuum $SU(2) \times SU(2)$ Higgsless model (either in flat space or AdS$_5$). Note
also that, for $a>1$, the squared ``decay-constant" $F^2_2$ is less than zero.

In section IV E we will further examine the relationship between four-site linear, three-site triangular, and HLS models.

\subsection{The gauge sector}

The gauge sector of the triangular moose model is determined by 
five parameters: $v$, $g$, $g'$, $\tilde{g}$, and $a$. Three of these
will be fixed by $G_F$, $\alpha_{em}$, and $M_Z$ -- leaving two
free parameters, which
we will take to be $a$ and the mass of the heavy charged boson $M_{W'}$.  We will describe the essential features of the gauge sector here and give details, such as the gauge boson mass-eigenstate wavefunctions,  in Appendix C.

The charged-boson mass matrix is given by
\begin{equation}
{\cal M}^2_W = 
\frac{\tilde{g}^2 v^2}{4}
\begin{pmatrix}
x^2(1+a) & -2xa \\
-2xa & 4a
\end{pmatrix}~,
\end{equation}
where $x=g/\tilde{g} \ll 1$. Expanding in $x$, we find
the masses
\begin{equation}
M^2_W = \frac{g^2 v^2}{4} \left(1-\frac{x^2}{4} + \frac{(a-1) x^4}{16 a} + \cdots\right)\,, \qquad \qquad
M^2_{W'}  = \tilde{g}^2 v^2 a \left(1+\frac{x^2}{4} + \frac{x^4}{16 a} + \cdots\right)~,
\label{eq:mw2}
\end{equation}
from which we may derive the relationship
\begin{equation}
x^2 = 4 a \left(\frac{M_W}{M_{W'}}\right)^2 + 8 a^2 \left(\frac{M_W}{M_{W'}}\right)^4
+ 4 a^2(5a+2) \left(\frac{M_W}{M_{W'}}\right)^6+\cdots~.
\label{eq:x}
\end{equation}

The neutral-boson mass matrix is given by
\begin{equation}
{\cal M}^2_Z = 
\frac{\tilde{g}^2 v^2}{4} 
\begin{pmatrix}
x^2(1+a) & -2xa & -t x^2 (1-a) \\
-2xa & 4a & -2txa\\
-t x^2 (1-a) & -2txa & t^2 x^2 (1+a)
\end{pmatrix}~,
\label{eq:neutralmat}
\end{equation}
where $t\equiv g'/g = s/c$ and $s^2 + c^2 = 1$. This matrix has a zero eigenvalue; its associated eigenvector corresponds to the photon, whose electric charge is related to the gauge couplings by
\begin{equation}
 \frac{1}{e^2} = \frac{1}{g^2} + \frac{1}{\tilde{g}^2} + \frac{1}{{g'}^2} \,.
\end{equation}
The light neutral gauge boson, which we associate with the $Z$, has a mass
\begin{equation}
 M^2_Z  = \frac{g^2 v^2}{4 c^2}
\left[ 1- \frac{x^2}{4} \frac{(c^2-s^2)^2}{c^2} + \frac{(a-1)(s^2-c^2)^2 x^4}{16 c^4 a} +\cdots\right] \,,
  \label{eq:ZmassHLS}
\end{equation}
while the mass of the heavy neutral $Z'$ boson is
\begin{equation}
 M^2_{Z'}  = \tilde{g}^2 v^2 a \left[1+ \frac{x^2}{4 c^2} + \frac{x^4(1-t^2)^2}{16 a} + \cdots\right]~.
 \label{eq:HLSzprmas}
\end{equation}
Note that all the expressions for quantities related to the gauge sector reduce to their counterparts in the three-site model if we set $a=1$.

\subsection{$Z$-standard weak mixing angle}

In order to compare the model with experiment, we will define a ``$Z$-standard" weak mixing angle $\sin\theta_Z \equiv s_Z$ in terms of $e$, $M_Z$ and $G_F$, as in Eq. (\ref{eq:zstdmx}).  This  is related to $s$, defined just below Eq. (\ref{eq:neutralmat}) through Eq. (\ref{eq:Zstandard}) where
\begin{equation}
 \Delta \equiv s^2 \left(c^2-\frac{1}{4}\right)x^2 + O(x^4).
\label{eq:DeltaHLS}
\end{equation}
In other words, $s^2$ and $s^2_Z$ differ only at order $x^2$.

We can also calculate the $ZWW$ and $WWWW$ vertices in the triangular moose model from the overlap of the gauge boson wavefunctions at the different sites.  Since the dependence of the wavefunctions on $a$ only starts at order $x^3$, the expressions for  $g_{ZWW}$ and $g_{WWWW}$ are the same as in the three-site model to order $x^2$:
\begin{eqnarray}
g_{ZWW} &=& \frac{e c_Z}{s_Z} \left[ 1 + \frac{1}{8 c^2} x^2 + {\cal O}(x^4)\right] = \frac{e c_Z}{s_Z} \left[ 1 + \frac{a}{2 c^2} \left(\frac{M_W}{M_W'}\right)^2 \right]\,, \label{eq:HLS_gzww}\\
g_{WWWW} &=& \frac{e^2}{s_Z^2} \left[ 1 + \frac{5}{16} x^2 + {\cal O}(x^4)\right]= \frac{e^2}{s_Z^2} \left[ 1 + \frac{5a}{4} \left(\frac{M_W}{M_W'}\right)^2 \right] \,,\label{eq:HLS_gzwww}
\end{eqnarray}
where the middle expressions above agree with those of the three site model found in Ref. \cite{SekharChivukula:2006cg}  and the right-hand ones result from applying Eq. (\ref{eq:x}).  The leading terms are precisely as in the Standard Model, and the size of the deviations in the couplings is proportional to $a$.   While the general form of the expressions resembles that for 5d $SU(2) \times SU(2)$ models, the comparison is complicated by the fact that the $ZWW$ and  $WWWW$ couplings in continuum models vary with geometry  \cite{Chivukula:2005ji}.

\subsection{The fermion sector}
\label{sec:fermiontriangle}

In the fermion sector of the model, we will consider a delocalized   \cite{SekharChivukula:2005xm,Cacciapaglia:2004rb,Foadi:2004ps}, massless fermion deriving its $SU(2)_W$ properties from both site 0 and site 1, just as in the three-site model.

To ensure that the model does not generate precision electroweak corrections of a size excluded by the data, and yet allow for the possibility of deviations from ideal delocalization \cite{SekharChivukula:2005xm}, we make the definition
\begin{equation}
g_L^{triangular}=g_L^{SM}\left(1+\frac{\alpha S_0}{4s_Z^2}\right)\,,
\end{equation}
which is related to the delocalization parameter $\epsilon_L$ as
\begin{equation}
\epsilon_L^2 = 
-\frac{\alpha S_0}{2s_Z^2}
+x^2\left(\frac{1}{2}-
\frac{(3a+1)\alpha S_0}{8as_Z^2}\right)
+\mathcal{O}\left(x^4,(\alpha S_0)^2\right)\,,
\label{eq:epLS}
\end{equation}
a relationship that reduces to the three-site model form for $a=1$.  In the special case of ideal delocalization, all four leading-order  parameters \cite{Barbieri:2004qk,Chivukula:2004af} that describe the deviations  of a flavor-universal extension of the Standard Model vanish and, in particular, the parameter  $\alpha S_0$  \cite{Peskin:1992sw,Altarelli:1990zd}
vanishes.  From (\ref{eq:epLS}), one may show that as one approaches ideal delocalization by sending $S_0 \to 0$, the corresponding value of $\epsilon_L$ becomes\footnote{In the 3-site-model limit, this choice of $\epsilon_L^2$ is equivalent to a choice of the parameter $b$ in \cite{Anichini:1994xx} to make $\epsilon_3$ or $\alpha S$ vanish.}
\begin{equation}
\epsilon_L^2 (S_0 \to 0) = \frac{x^2}{2} + \frac{x^4}{8 a^2} + \cdots\,,
\end{equation}
which reproduces the three-site model result for a=1.

To see that the principle of ideal delocalization applies for any value of $a$, one can examine the form of the couplings between gauge bosons and ideally delocalized fermions  when written in terms of the physical quantities $e$, $M_Z$ and $M_W$.  These may be obtained by computing the superposition of the eigenstate wavefunctions with the coupling strength at each site. The resultant expressions for the coupling of the left-handed fermion to the $W$ boson ($g_{L}^{W}$), the $T_3$-coupling of the left-handed fermion to the $Z$ boson ($g_{3L}^{Z}$), and the hypercharge coupling of the left-handed fermion to the 
$Z$ boson ($g_{YL}^{Z}$) are 
\begin{equation}
 g_{L}^{W} = \frac{e}{\sqrt{1-\frac{M_W^2}{M_Z^2}}} 
     \left[ 1 + O(x^4) \right] \,, \qquad
 g_{3L}^{Z} = \frac{e M_W}{M_Z\sqrt{1-\frac{M_W^2}{M_Z^2}}} 
     \left[ 1 + O(x^4) \right] \,, \qquad
 g_{YL}^{Z} = -\frac{e M_Z}{M_W}\sqrt{1-\frac{M_W^2}{M_Z^2}} 
     \left[ 1 + O(x^4) \right] .
     \label{eq:HLSferms}
\end{equation}
Clearly, the fermion couplings to the $W$ and $Z$ in the triangular Higgsless model are of very nearly the Standard Model form, as consistent with ideal fermion delocalization, for any value of $a$.

\subsection{$g_{Z'WW}$ value and comparison to continuum models}
\label{sec:HLSheavyvert}

We can calculate the $Z'WW$ and $ZW'W$ couplings in the triangular moose model by accounting for the wavefunction overlap among the gauge boson mass eigenstates at each site.   Using Eqs. (\ref{eq:x}), (\ref{eq:wvec}), (\ref{eq:wpvec}),   (\ref{eq:triang-z}), and (\ref{eq:triang-zp}) we find the following:
\begin{eqnarray}
g_{Z'WW} &=& g\, v^0_{Z'}(v^0_W)^2  + \tilde{g}\, v^1_{Z'}(v^1_W)^2 =  -\frac{\sqrt{a}}{2}\,\frac{e}{s_Z} \left(\frac{M_W}{M_{W'}}\right)  + \cdots \,,
\label{eq:Z'WW_HLS}\\
g_{ZW'W} &=& g\, v^0_{Z}v^0_{W'}v^0_W + \tilde{g}\, v^1_{Z}v^1_{W'}v^1_W =  -\frac{\sqrt{a}}{2}\,\frac{e}{s_Z c_Z} \left(\frac{M_W}{M_{W'}}\right) + \cdots .
\label{eq:ZW'W_HLS}
\end{eqnarray}
From these results, we can see that the triple-gauge-boson couplings involving a heavy gauge boson are suppressed by a factor of $M_W/M_{W'}$ compared to the magnitude of the Standard Model triple-gauge-boson coupling, $g_{ZWW} \simeq e\,c_Z/s_Z$.   We also note that  the contributions to these couplings from site 0 and site 1 enter at  the same order. Thus, we expect, in the continuum theory, that the contributions to $g_{Z'WW}$ and $g_{ZW'W}$ from triple gauge boson couplings in the bulk and on the 
brane are also comparable. This is consistent with our earlier observation that it was necessary to calculate the wave function of the heavy gauge 
bosons to the next-to-leading order to obtain the leading  contribution to $g_{Z'WW}$ and $g_{ZW'W}$ for the $SU(2) \times SU(2)$ continuum model studied in Sec. \ref{sec:twoa}.

To evaluate the triangular moose model as an effective theory for the $SU(2) \times SU(2)$ continuum models, let us first calculate the gauge coupling ratios of Eqs.~(\ref{eq:firstratio}) and (\ref{eq:secondratio}), using our expressions for the triple and quartic gauge couplings in Eqs.~(\ref{eq:HLS_gzww}), (\ref{eq:HLS_gzwww}), (\ref{eq:Z'WW_HLS}), and  (\ref{eq:ZW'W_HLS}).  We find that the first ratio has the value 1 in the triangular moose model, just as in the three-site and continuum  $SU(2) \times SU(2)$ models.  The second ratio has the value $\frac34 a$, in the triangular moose model, as compared with 1.0 in the continuum models and $\frac34$ in the three-site model; this suggests that an triangular moose model with $a = \frac43$ may provide a better approximate description of the continuum $SU(2) \times SU(2)$ models than the three-site model.

In fact, comparing Eqs. (\ref{eq:thirdahat}) and (\ref{eq:Z'WW_HLS}), while recalling that there is only one set of KK bosons in the triangular Higgsless model and that the $Z'$ and $W'$ are essentially degenerate as in Eqs. (\ref{eq:mw2}) and (\ref{eq:HLSzprmas}), reveals that the parameter $a$ in the triangular moose model is precisely the quantity $\hat{a}$ defined earlier as a means of comparing continuum and deconstructed models:
\begin{equation}
\hat{a}^{triangular} = \frac{v^2 g_{Z'WW}^2 M_{Z'}^2}{M_W^4} \simeq a~.
\end{equation}
This may also be confirmed by calculating the coupling strength of the $V\pi\pi$ vertex from Eq.~(\ref{eq:HLS}), which leads to the relation
\begin{equation}
 \hat{a}^{triangular}  = \frac{4 v^2 g_{V\pi\pi}^2}{M_V^2} \simeq a
\label{eq:couplings}
\end{equation}
where the LHS follows from Eq. (\ref{eq;fourthahat}).  Similarly, if we evaluate 
 $g_{\pi\pi\pi\pi}$ using the HLS Lagrangian (\ref{eq:HLS}), we find that it has the form
\begin{equation}
  g_{\pi\pi\pi\pi} = 1 - \frac34 a \,.
\end{equation}
 The 4-pion contact interaction vanishes \cite{Harada:2003jx,Harada:2006di} for $a = 4/3$ , reducing the sum rule (\ref{eq:E2sumd}) for this deconstructed model to the form of its 5d counterpart (\ref{eq:E2sum}).  
 
\begin{figure}
  \begin{center}
    \includegraphics[width=0.60\textwidth]{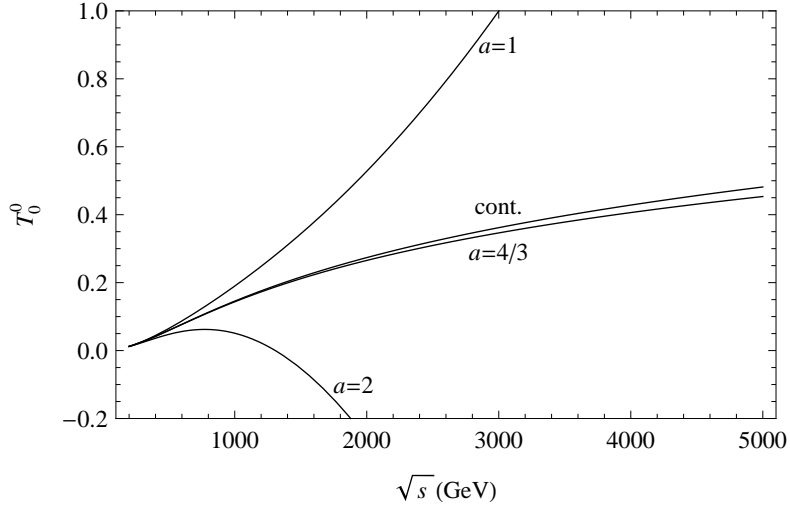}
  \end{center}
\caption{Behavior of the partial wave amplitude $T_0^0$ for pion-pion scattering in the triangular moose model with 
various $a$. The values $v=250$ GeV, $M_V=500$ GeV are assumed. The curve labeled ``cont." shows 
$T_0^0$ in the continuum flat $SU(2)\times SU(2)$ model for $M_1 = 500$ GeV.}
\label{fig:t00}
\end{figure}

The equivalence of $a$ and $\hat{a}^{triangular}$, allows us to make an interesting comparison of the amplitude for pion-pion scattering, which depends both on the four-pion contact interaction and on the $V\pi\pi$ vertex, in various models.   Fig.~\ref{fig:t00} shows the partial wave  amplitude $T_0^0$ for pion-pion scattering in the global\footnote{That is, we set $g = g' = 0$ for simplicity.} continuum flat $SU(2)\times SU(2)$ model with $M_1=500$ GeV compared with $T_0^0$ in the hidden local symmetry model for several values of the parameter $a$. 
The result in the global$^8$ three-site model is shown by the curve labeled $a=1$; the value $a=2$ is motivated by the phenomenological KSRF relation \cite{Kawarabayashi:1966kd,Riazuddin:1966sw}.
This plot ties our results together quite neatly: while the curves with three different values of $a$ all give a reasonable description of $T_0^0$ at very low energies (as dictated by the low-energy theorem \cite{Weinberg:1966kf,Chanowitz:1986hu,Chanowitz:1987vj}), the best approximation to the continuum behavior of $T_0^0$ over a wide range of energies is given by the triangular moose model curve with $a=4/3$.   At low energies, the fact that the three-site and  triangular models both prevent $E^4$ growth of the amplitude suffices; but at higher energies, the fact that the triangular moose model has $g_{\pi\pi\pi\pi} = 0$ and $\hat{a} = 4/3$ enables it to cut off the $E^2$ growth of the amplitude as well, as consistent with the behavior in the continuum model.

\subsection{Reduction of a four-site model to the triangular moose}
\label{sec:foursite}

\begin{figure}
\begin{center}
\includegraphics[width=8cm]{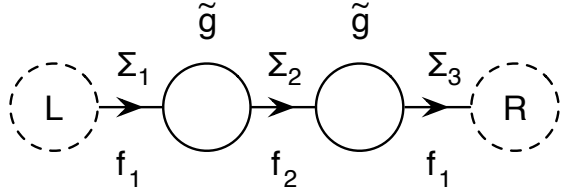}
\end{center}
\caption{The four-site global moose diagram for the model considered in
Section \protect\ref{sec:foursite}. 
The model has an $SU(2)_L \times SU(2)_R$ global symmetry and two gauged
$SU(2)$ groups  with equal couplings, 
and therefore has a left-right parity symmetry. We
consider the ``reduction" of this model in the limit $f_2 \to \infty$.
\label{fig:one}}
\end{figure}

\begin{figure}
\begin{center}
\includegraphics[width=6cm]{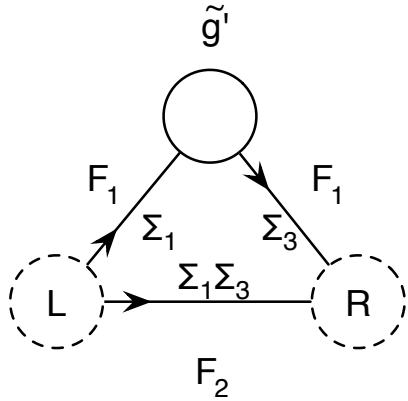}
\end{center}
\caption{Triangular moose model to which the four-site moose in Fig. \ref{fig:one} 
reduces in the limit $f_2 \to \infty$.
Note that the three link fields are not independent, and the ``non-local" link sigma-model
field is $\Sigma_1 \cdot \Sigma_3$.
We determine the values of $F^2_1$ and $F^2_2$  that result, and find that $F^2_2 < 0$.
\label{fig:two}}
\end{figure}

In the previous subsection, we have demonstrated that the HLS or BESS triangular Higgsless model
with $a=4/3$ accurately describes the properties of $W_L W_L$ elastic scattering in continuum Higgsless models. One puzzling aspect of this conclusion is that the triangular moose model is non-local
in ``theory space" \cite{ArkaniHamed:2001ca}, whereas the most general continuum Higgsless
model can be derived \cite{SekharChivukula:2008mj} as a limit of the linear Higgsless model illustrated in Fig. \ref{fig:generalmoose}, which is local in theory space. In this subsection we
show how the triangular moose model arises naturally from a linear moose model when one ``integrates out" heavy vector modes; we illustrate how this occurs by showing how the four-site global linear model illustrated in Fig. \ref{fig:one} reduces to to a three-site triangular moose model illustrated in Fig. \ref{fig:two}  in the limit in which $f_2 \to \infty$.  Note that the four-site moose in this case is not our ``minimal" $SU(2)^2 \times U(1)^2$ moose, but the alternative $SU(2)^3 \times U(1)$ moose.

Let us begin by examining the global four-site model.
Given the equal couplings of the two gauged $SU(2)$ groups, the
model has a parity symmetry under which 
\begin{align}
\Sigma_1 \leftrightarrow \Sigma^\dagger_3\,, \nonumber\qquad\qquad  
\Sigma_2  \leftrightarrow \Sigma^\dagger_2\,, \qquad\qquad
W^\mu_1  \leftrightarrow W^\mu_2~,\nonumber
\end{align}
where $W^\mu_{1,2}$ are the gauge bosons of the two gauge-groups.
In addition, we can also see by inspection that 
the decay constant for the exact
Nambu-Goldstone bosons of the four-site model (corresponding to the
electroweak scale) is related to the $f$-constants of the individual links as
\begin{equation}
\frac{1}{F^2_\pi} = \frac{2}{f^2_1} + \frac{1}{f^2_2} = \frac{1}{v^2}~.
\end{equation}
As consistent with Georgi's spring analogy \cite{Georgi:2004iy}, the $f$-constants for the three
links in linear progression from group L to group R combine like the spring-constants 
of three springs in series.

Working in a ``unitary'' gauge in which $\Sigma_2 \equiv {\cal I}$, we may write
the Lagrangian for the four-site model to leading order as
\begin{equation}
{\cal L} =  -\,\frac{1}{4 \tilde{g}^2}\left[(W^{\mu\nu}_1)^2 + (W^{\mu\nu}_2)^2\right] 
+\frac{f^2_2}{4} {\rm Tr}(W^\mu_1 - W^\mu_2)^2 
+ \frac{f^2_1}{4} {\rm Tr}\left[D^\mu \Sigma^\dagger_1 D_\mu \Sigma_1
+ D^\mu \Sigma^\dagger_3 D_\mu \Sigma_3\right]~,
\label{eq:4sitelag}
\end{equation}
Here,  the fields $W^\mu_{1,2} = W^{\mu a}_{1,2}\cdot  \tau^a$
are Hermitian matrix vector fields where the $\tau^a$ are the generators of $SU(2)$,
and the covariant derivatives are given by
\begin{align}
D^\mu \Sigma_1 & = \partial^\mu \Sigma_1 - i \Sigma_1 W^\mu_1~,\\
D^\mu \Sigma_2 & = \partial^\mu \Sigma_2 + i W^\mu_1 \Sigma_2 - i \Sigma_2 W^\mu_2 \to
i(W^\mu_1 - W^\mu_2)~,\\
D^\mu \Sigma_3 & = \partial^\mu \Sigma_3 + i W^\mu_2 \Sigma_3~.
\end{align}

Our strategy is to rewrite the $f$-constants in the form
\begin{equation}
f_1 = \frac{\sqrt{2} v}{\cos\alpha}~, \ \ \ \ f_2 = \frac{v}{\sin\alpha}~,
\end{equation}
and consider the limit $\sin\alpha \to 0$ with $v$ fixed. 
In this limit, we see from Eq. (\ref{eq:4sitelag}) that
the ``axial" vector field proportional to $W^\mu_1 - W^\mu_2$ becomes heavy.
Following \cite{Sekhar Chivukula:2006we} we will integrate out the heavy field
at tree-level by finding the equation of motion
for the axial field, solving this order-by-order in $\alpha$, and plugging the solution back
into the Lagrangian in Eq. (\ref{eq:4sitelag}). 

The equations of motion of $W^\mu_{1,2}$ can be written \cite{Sekhar Chivukula:2006we} in the form 
\begin{align}
W^\mu_1 = & \frac{1}{f^2_1+f^2_2}\left[
f^2_2 W^\mu_2 - i f^2_1 \Sigma^\dagger_1 \partial^\mu \Sigma_1\right] + \cdots ~,\\
W^\mu_2  =  & \frac{1}{f^2_1+f^2_2}\left[
f^2_2 W^\mu_1 - i f^2_1 \Sigma_3 \partial^\mu \Sigma^\dagger_3\right] + \cdots ~,
\end{align}
in the gauge in which $\Sigma_2 \equiv {\cal I}$ and neglecting terms with higher derivatives.
Taking the difference between these two equations, we find
\begin{equation}
W^\mu_1 - W^\mu_2 = -i\,\sin^2\alpha\,\left(\Sigma^\dagger_1 \partial^\mu \Sigma_1
- \Sigma_3 \partial^\mu \Sigma^\dagger_3\right) + \cdots~.
\label{eq:axial}
\end{equation}
In order to use this expression to study the $\alpha \to 0$ limit of
the Lagrangian of Eq. (\ref{eq:4sitelag}), it is most convenient to rewrite the
gauge fields in the notation
\begin{equation}
W^\mu_{1,2} = W^\mu \pm \frac{1}{2}(W^\mu_1 - W^\mu_2)~, {\rm\ \  where\ \ }  W^\mu \equiv \frac{1}{2} (W^\mu_1 + W^\mu_2)~,
\label{eq:Wdecompose}
\end{equation}
The newly-defined $W^\mu$ is the appropriate ``block-spin" vector field resulting from integrating
out the middle link, and also corresponds to the physical light vector boson state. We may, likewise, use Eq. (\ref{eq:Wdecompose}) to define the  covariant derivative $\tilde{D}^\mu$:
\begin{eqnarray}
\label{eq:fromhere}
\tilde{D}^\mu \Sigma_1 &= & \partial_\mu \Sigma_1 -i\,\Sigma_1 W^\mu~, \\
\tilde{D}^\mu \Sigma_3 &= & \partial_\mu \Sigma_3 + i\,W^\mu \Sigma_3~,
\end{eqnarray}
and then rewrite several key expressions appearing in the four-site Lagrangian (\ref{eq:4sitelag}) as follows:
\begin{eqnarray}
\Sigma^\dagger_1 D^\mu \Sigma_1 &=  & \left(1-\frac{\sin^2\alpha}{2}\right)
(\Sigma^\dagger_1 \tilde{D}^\mu \Sigma_1) + \frac{\sin^2\alpha}{2}
\Sigma_3 \tilde{D}^\mu \Sigma^\dagger_3 + \cdots~, \\
(D^\mu \Sigma_3)\Sigma^\dagger_3 &=  &\left(1-\frac{\sin^2\alpha}{2}\right)
(\tilde{D}^\mu \Sigma_3)\Sigma^\dagger_3 + \frac{\sin^2\alpha}{2}
(\tilde{D}^\mu\Sigma^\dagger_1)\Sigma_1 + \cdots~,\\
W^\mu_1 - W^\mu_2 &= & -i\,\sin^2\alpha\,\left( 
\Sigma^\dagger_1 \tilde{D}^\mu \Sigma_1 - \Sigma_3 \tilde{D}^\mu \Sigma^\dagger_3
\right) + \cdots~.
\label{eq:tohere}
\end{eqnarray}

Applying Eqs. (\ref{eq:fromhere}) -- (\ref{eq:tohere}) to the four-site Lagrangian (\ref{eq:4sitelag}), we find that it takes the following form at ${\cal O}(p^2)$ 
\begin{equation}
{\cal L}^{triangular}_{p^2} = \frac{2 v^2 }{4 \cos^2\alpha} {\rm Tr} \left[ (\tilde{D}_\mu \Sigma_1)^\dagger(\tilde{D}^\mu \Sigma_1)  +   (\tilde{D}_\mu \Sigma_3)^\dagger(\tilde{D}^\mu \Sigma_3) \right]  - \frac{v^2 \sin^2\alpha}{4 \cos^2\alpha}\,  {\rm Tr} \left[ (\tilde{D}_\mu (\Sigma_1\Sigma_3))^\dagger  (\tilde{D}_\mu (\Sigma_1 \Sigma_3))  \right]\,.
\label{eq:newtriangle}
\end{equation}
This reduced Lagrangian corresponds 
to the triangular moose (\ref{eq:triangle-Lag}) illustrated in Fig. \ref{fig:two} with the identifications 
\begin{equation}
F^2_1 =  \frac{2 v^2}{\cos^2\alpha} ~,\qquad\qquad 
F^2_2 =  -\frac{v^2\sin^2\alpha}{\cos^2\alpha}~.
\end{equation}
This time, Georgi's spring analogy \cite{Georgi:2004iy}, leads us to expect that the $f$-constants 
for the links from group L to group R will combine as for the spring constants of two
identical springs in series with one another and jointly in parallel with a third
spring.  And this is exactly what we find:
\begin{equation} 
F^2_\pi = \frac{F^2_1}{2}+F^2_2 = v^2~.
\end{equation}
Note that the reduced Lagrangian is non-local in theory space, and includes 
a negative $F^2$.

We may also compute the effect of the decomposition in Eq. (\ref{eq:Wdecompose})
on the gauge-boson kinetic energy terms. Since $W^\mu_1 - W^\mu_2 ={\cal O}(\sin^2\alpha)$ in the limit of small $\sin\alpha$,
we find that the only effect is to produce a coupling constant for the block-spin gauge-boson
$W^\mu$ equal to $\tilde{g}/\sqrt{2}$. Hence, we can show that  the vector-meson mass-squared $M^2_{W'}$ 
and coupling constant $g_{W'\pi\pi}$
also agree to ${\cal O}(\sin^4\alpha)$ between the four-site and reduced
Lagrangians.

Finally, as in Section IV, we recall that the triangular moose Lagrangian (\ref{eq:triangle-Lag}) is equivalent to the HLS Lagrangian \cite{Bando:1985ej,Bando:1985rf} shown in Eq. (\ref{eq:HLS}) if we establish the correspondence shown previously in Eqs. (\ref{eq:tri-HLS-corresp}):
\begin{align}
F^2_1 & = 2 a v^2 ~,\nonumber \\
F^2_2 & = v^2(1-a)~. \nonumber
\label{eq:tri-HLS-corresp}
\end{align}
Thus, we conclude that the $[SU(2)]^3\times U(1)$ four-site moose in the limit of small $\sin\alpha$ corresponds to an HLS model with $a \approx 1$:
\begin{equation}
a =\frac{1}{\cos^2\alpha}= 1 + \sin^2\alpha + \cdots\,.
\end{equation}
%

\section{Phenomenology}

We now look at the collider phenomenology of the models discussed in this paper.  First, we update the LEP II bounds on $M_W'$ in these models.  Second, we discuss searches for a $W'$ or $Z'$ boson at the  LHC in channels that depend only on the $ZW'W$ or $Z'WW$ couplings.   Those heavy-light-light triple gauge boson vertices have a common value in continuum theories of different geometries, as in Eq. (\ref{eq:similarities}), and have similar values even in the various effective theories (three-site, four-site, triangular) discussed here.  Hence this $W'$ and $Z'$ phenomenology will be characteristic of this whole suite of theories.  For a given gauge boson mass, one can make a robust prediction of the size of the signal expected for the entire class of theories -- a good situation for an initial discovery search at the LHC.  Third, we consider the rather different scenario presented by measurements of the $Z'$ line shape or  $ZWW$ vertex at future high-energy lepton colliders.  We will see that the experiments should be quite sensitive to the value of $a$ in HLS-type models, and to the $Z'$ mass and $Z'WW$ in general.  A precise measurement at the ILC could thus potentially discriminate among various continuum and deconstructed theories -- a good situation for a follow-on measurement to an LHC discovery of a $W'$ or $Z'$.

\subsection{LEP}
\label{sec:LEP}

Studies of $WW$ production at LEP-II provide important constraints on deconstructed Higgsless models with nearly-ideal delocalization.  In this section, we extend an analysis \cite{Chivukula:2005ji, SekharChivukula:2006cg} that was previously undertaken for the three-site model where the fermions were taken to be ideally delocalized and the only modification considered was a change to the $ZWW$ coupling.  We find that a more general analysis allowing for deviations from ideal delocalization and incorporating both $s$-channel and $t$-channel $WW$ production
sub-processes gives a more complete picture. In this work we also incorporate the effects of
the extra link in the triangular Higgsless model.

We begin by reviewing the Hagiwara-Peccei-Zeppenfeld-Hikasa triple-gauge-vertex formalism \cite{Hagiwara:1986vm}, which gives the most general CP-invariant form of the $ZWW$ vertex as
\begin{equation}
{\cal L}_{ZWW}  =  -ie\frac{c_Z }{s_Z}\left[1+\Delta\kappa_Z\right] W^+_\mu W^-_\nu Z^{\mu\nu}
- i e \frac{c_Z}{s_Z} \left[ 1+\Delta g^Z_1\right](W^{+\mu\nu}W^-_\mu - W^{-\mu\nu} W^+_\mu)Z_\nu 
\label{eq:tgvlag}
\end{equation}
where the two-index tensors denote the Lorentz field-strength tensors of the corresponding fields.   In the Standard Model, $\Delta\kappa_Z = \Delta g^Z_1 \equiv 0$.   In any vector-resonance model, such as the Higgsless models considered here, to lowest order the interactions (\ref{eq:tgvlag}) come from re-expressing the nonabelian couplings in the kinetic energy terms of the original Lagrangian in terms of the mass-eignestate fields. Here, this yields equal contributions to the two anomalous couplings: $\Delta g^Z_1 = \Delta \kappa_Z$ \cite{Chivukula:2005ji}. The parameter $\Delta g^Z_1$ has been bounded by LEP-II  \cite{LEPEWWG} as $-.054< \Delta g^Z_1 < 0.028$ at 95\% CL.  Note that this formalism only addresses changes to the $s$-channel diagrams through alterations of triple-gauge couplings and does not incorporate the possibility of changes to $t$-channel neutrino-exchange diagrams via alterations to the $We\nu$ coupling or the presence of a new vector-like neutrino state.  
 
Because the Higgsless models considered here include physics that could change $t$-channel contributions to $WW$ production at lepton colliders, we have undertaken a pair of calculations in order to understand how the LEP-II bound on $\Delta g^Z_1$ applies.  First, we used CalcHEP 
\cite{Pukhov:2004ca} to calculate the forward ($\cos\theta > 0.75$) scattering cross-section\footnote{We also calculated the total scattering cross-section and various asymmetries, but found the forward cross-section to be the most sensitive to deviations from the Standard Model.} for $e^+ e^- \to W^+ W^-$ in the case where the Standard Model Lagrangian is modified by inclusion of the new physics terms in (\ref{eq:tgvlag}).  This enabled us to make contact with the procedure followed by the LEP experiments. Then we separately implemented our Higgsless model in  CalcHEP using the FeynRules \cite{Christensen:2008py} package and calculated the forward scattering cross-section as a function of $M_{W'}$ for various values of the parameters $a$ and $S_0$ (including the case of ideal delocalization where $S_0$ is close to zero).  

In both of our calculations, we included all relevant $s$-channel and $t$-channel diagrams.\footnote{The heavy fermion masses in the Higgsless model were taken to be 4 TeV, though
any acceptable heavy fermion mass \protect\cite{SekharChivukula:2006cg} above 2 TeV yields similar results.}  We find that the pure $s$-channel, pure $t$-channel and interference terms in the cross section each display similar sensitivity to changes in the value of $M_{W'}$ and/or $\Delta g_1^Z$ and that there is an important, partial cancellation among them.  As a result, using the LEP-II limit on $\Delta g_1^Z$ to bound $M_{W'}$ in Higgsless models requires that both $s$-channel and $t$-channel diagrams be included in a full analysis\footnote{Even in a case where the pure $t$-channel piece of the cross-section has no new-physics contribution (e.g. if one sets $S_0 = 0$ so that the $We\nu$ coupling is of purely standard form), the interference between the s- and $t$-channel amplitudes is still significant for understanding how the cross-section varies with $M_{W'}$ or $\Delta g^Z_1$. }.  

\begin{figure}
\begin{center}
\includegraphics[scale=1.0]{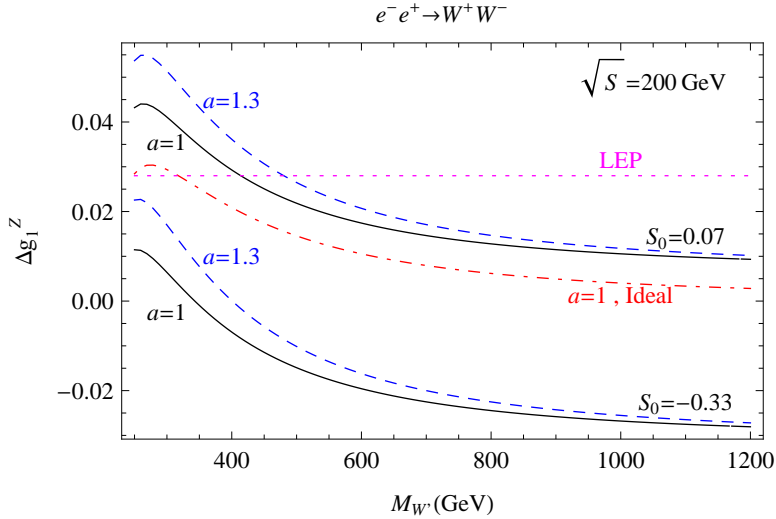}
\end{center}
\caption{\label{fig:LEP-II} The effective $\Delta g_1^Z$ in the triangular moose model.  This effective value was computed by calculating the forward cross section in the Standard Model with $\Delta g_1^Z$ incorporated and separately calculating the forward cross section in the triangular  model.  The cross sections were then compared and the value of $\Delta g_1^Z$ that gave the same effect as in the triangular moose model was determined.  The value of this parameter for the three-site model is obtained by setting $a=1$.  The curve labeled ``a=1,Ideal" corresponds to the three-site model in the limit of ideal delocalization.  The horizontal line labeled ``LEP" indicates the LEP-II upper bound 
\protect\cite{LEPEWWG} of $\Delta g_1^Z\leq0.028$ at 95\% CL.}
\end{figure}

By comparing our two calculations, we find the effective value of $\Delta g^Z_1$ that gives the same forward cross-section as our deconstructed Higgsless models for various values of $M_{W'}$, $a$, and $S_0$, and we present the results, together with the LEP-II upper bound on $\Delta g^Z_1$, in figure \ref{fig:LEP-II}.  As this plot shows, increasing the value of $S_0$ or $a$ increases the lower bound on $M_{W'}$, while lowering them decreases the lower bound on $M_{W'}$.  For some values of $S_0$ and $a$, the corresponding curve never rises above the LEP-II upper bound on $\Delta g^Z_1$; however, as $M_{W'}$ (and therefore $M_{Z'}$) is reduced further, each curve reaches a maximum and quickly drops below the LEP-II upper bound on $\Delta g^Z_1$.   As a result, it is not expected that the lower bound on $M_{W'}$ will be much below $250$ GeV for any values of the model parameters. Note that for the case of models with $a=1$ and ideal delocalization, the lower bound on $M_{W'}$ is about 320 GeV.   This is slightly less-restrictive than the 3-site model bound in \cite{Chivukula:2005ji, SekharChivukula:2006cg}, which was obtained by considering only $s$-channel processes.\footnote{If we include only the $s$-channel diagrams in our present calculation, we reproduce the results of \cite{Chivukula:2005ji, SekharChivukula:2006cg}}

\subsection{LHC}

LHC searches for a $W'$ boson have the potential to confirm or constrain the Higgsless theories discussed here.  As shown in  \cite{He:2007ge}, the fermiophobic $W'$ boson of the three-site
 model with ideal delocalization can be produced at the LHC through either the vector boson fusion process  $pp \to W' jj$ or the associated production process $pp \to W' Z$; in either case, the $W'$ then decays dominantly to $WZ$. A set of simple cuts can render the $W'$ signal visible at the LHC in both processes for its entire allowed mass range, roughly 400 to 1200 GeV  \cite{He:2007ge}.  To give an example, for $M_{W'}=500\,(400)$\,GeV, the $5\sigma$ discovery of $W'$ requires an integrated luminosity  of  12\,(7)\,fb$^{-1}$ for the vector boson fusion process $pp \to  W' jj \to W Z  jj \to \nu3\ell\,jj$ and an integrated luminosity of  26\,(7.8)\,fb$^{-1}$  for the associated production process $pp \to W'Z \to W Z Z \to jj\,4\ell$.

Because the $W'$ boson of our four-site model is essentially identical to that of the $W'$ in the 3-site model, its hadron collider phenomenology is the same.

The $W'$ of the triangular moose model for general values of $a$ would also have the same production and decay modes at the LHC as the $W'$ boson of the three-site model, but would not necessarily look identical to it.  The main difference is that the $ZW'W$ vertex involved in the production of the $W'$ in either process would be (for a given $M_{W'}$ mass) proportional to $\sqrt{a}$, as discussed above in Sec. \ref{sec:HLSheavyvert}, while the produced $W'$ will still decay to $WZ$ with a branching fraction of nearly 100\% .  Hence the signal cross-section will be proportional to $a$, while the background will be unchanged.  The integrated luminosity required for discovery at a given confidence level will, therefore, be reduced if the value of $a$ exceeds 1 (its value in the three-site model). For example, as a rough estimate, the luminosity required for a 3-sigma signal in the three-site model would provide a $5\sigma$ signal in the triangular moose model with $a \approx \sqrt{3}$, to judge by the curves in Fig. 4 of \cite{He:2007ge}.

A similar pattern should hold for the contribution of $Z'$ bosons to $WW$ scattering or $Z'W$ associated production in these models: the four-site signals will resemble those of the three-site model, while the rate expected for an triangular moose model $Z'$ boson can differ due to the factor of $\sqrt{a}$ in the $Z'WW$ coupling.  A caveat is in order: discovering a $Z'$ boson in either vector boson fusion ($pp \to Z' jj \to WW jj$) or associated production ($pp \to Z' W \to WWW$) would be difficult\footnote{A recent parton-level study \cite{Ohl:2008ri} of the three-site model, including the possibility of other-than-ideal delocalization, suggests that a $Z'$ produced by Drell-Yan $q\bar{q}$ annihilation and decaying to $W$ pairs may be visible at the LHC; however, the effects of hadronization, detector response, and backgrounds from $WW$ fusion remain to be explored.} due to the difficulty in reconstructing the $Z'$ peak in leptonic $W$-decays and
the large backgrounds in hadronic $W$-decays.  These signals are most likely to be useful for confirming the presence of a $Z'$ state whose existence and mass are suggested by the prior discovery of a $W'$ boson.  

More specifically,  in the $SU(2)^2 \times U(1)^2$ 4-site model with $u^2 \tilde{q}^2 \left[f'/f\right]^2 =2$,  the $Z'$ and $Z''$ bosons are almost degenerate (as is approximately true in continuum
$SU(2) \times SU(2) \times U(1)$ Higgsless models, as discussed in Sec. \ref{sec:continuum}), and, in $s$-channel scattering processes, they would most likely appear as a single resonance.  For instance, if the mass of the $Z'$ boson were 500 GeV, we calculate that the mass difference between the $Z'$ and $Z''$ would be $\sim$14 GeV and the width of each boson would be a few GeV.  Given the anticipated 10-15 GeV dijet energy resolution at the LHC  \cite{ATLAS}, it is unlikely that the $Z'$ and $Z''$ would be separately distinguishable; of course, improved detector resolution could ultimately alter this conclusion. 
The combined tree level contributions of $Z'$ and $Z''$ exchange to WW scattering or $Z'W$ associated production will, then, look like the exchange of a single heavy gauge boson with coupling strength $g_{Z'WW}^2 + g_{Z''WW}^2$, which is equal to the value of $g_{Z'WW}^2$ in the three-site model.  Hence the single $Z'$ boson of the three-site model should closely approximate (at leading order) the joint contribution of $Z'$ and $Z''$ exchange in the minimal 4-site model with $u^2 = 2$.  On the other hand, in the limit where $u^2\tilde{q}^2 \left[f'/f\right]^2 - 2 = O(1)$, we have seen that the $Z'$ boson has the same coupling to $WW$ as the $Z'$ of the three-site model, while the $Z''WW$ coupling is suppressed by an additional factor of $x^2$.  In this limit, it is just the $Z'$ by itself that could visibly affect $WW$ scattering, and its effect would be the same size as that in the three-site model.

\subsection{ILC}

\begin{figure}[ht]
\begin{center}
\includegraphics[angle=0,height=0.45\textwidth,width=0.60\textwidth]{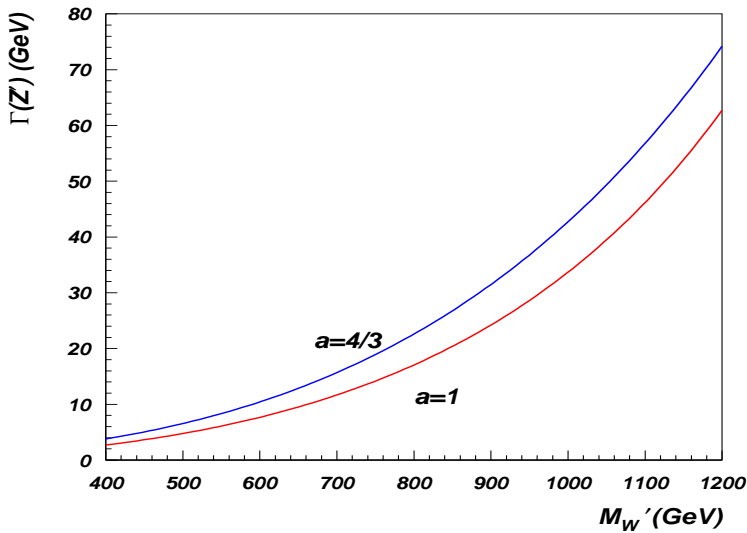}
\caption{Width of the $Z'$ boson as a function of $M_{W'}(=M_{Z'})$ in the triangular moose model for
$a=1$ (equivalent to the three site model) and $a=4/3$.
\label{fig:width_vs_mwp}
}
\end{center}
\end{figure}

The primary aim of an ILC study of Higgsless models
would be a precise determination of the underlying model parameters.
In particular, in case of the triangular moose model
the precision measurement of the $W'$ mass and the $a$-parameter
are crucial. Given that the  $Z'$ and $W'$ should not weigh much more than a TeV and given
the projected ILC centre-of-mass energy of up to 1 TeV \cite{Brau:2007zza},
the $W'$ and $Z'$ could be kinematically accessible at the ILC.
If direct $Z'$ production is possible,
the mass as well as the  width could be precisely measured via 
a $Z'$ shape-line study, in analogy to the LEP-I
$Z$ boson measurements.
In Fig.~\ref{fig:width_vs_mwp}, we plot
$\Gamma_{Z'}$ versus $M_{W'}$, and 
we see that an accurate determination
of the width (for a given mass) will allow a precise determination
of the parameter $a$.

\begin{figure}[htb]
\begin{center}
\includegraphics[angle=0,height=0.45\textwidth,width=0.60\textwidth]{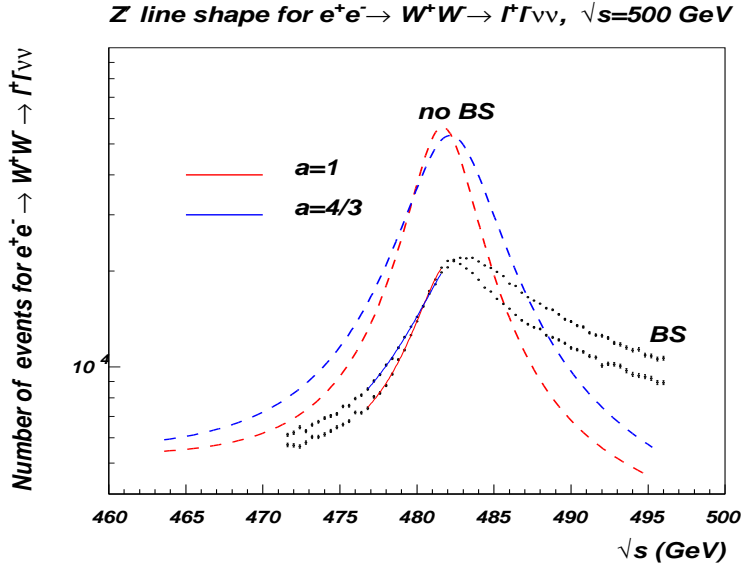}
\caption{
\label{fig:width_fit}
The $Z'$ width
measurement from an energy-scan at the ILC.
The dashed lines show the $Z'$ fitted  line shapes  
without  ISR or BS taken into account,
with $M_{W'}=480$~GeV
from the $e^+e^- \to W^+W^- \to \ell^+\ell^-\nu\nu$
process event rate  for $a=1$(red) and $a=4/3$(blue).
Small open circles with vertical error bars represent   ``data" simulated by
taking into account  ISR and BS effects.}
\end{center}
\end{figure}

In Fig. \ref{fig:width_fit}
we present the results of a simulation of a future $Z'$ width
measurement at the ILC.
The dashed lines present the fitted $Z'$  line shapes 
(with $M_{W'}=480$~GeV)
using the $e^+e^- \to W^+W^- \to \ell^+\ell^-\nu\nu$
process by counting the number of
final state events for $a=1$ (red) and $a=4/3$ (blue).
These dashed line shapes have been obtained without taking 
Initial State Radiation (ISR)~\cite{Kuraev:1985hb,
Jadach:1988gb,Skrzypek:1990qs} 
and Beamstrahlung (BS) effects~\cite{Chen:1991wd} into account.  They 
should  therefore be considered as  ``ideal" line shapes, 
 which would be inferred   from the real
line shapes shown by the small open circles of  ``data" 
(denoted by ``BS" in the Fig. \ref{fig:width_fit}) which were simulated by
taking into account  ISR and BS effects as implemented in CalcHEP
\cite{Pukhov:2004ca}.\footnote{
We used the following set of parameters defining the  Beamstrahlung:
horizontal beam  size = 640 nm,
vertical   beam  size = 5.7 nm, 
bunch length = 0.3mm ,
number of particles per bunch= $2\times10^{10}$~.
}
The ISR and BS  effects lead to a shift in the position of the maximum of the line
shape by  about $0.5$~GeV (i.e. about a +1\% shift), as well as a distortion of the
line shape. This simulated effect is
in good  agreement with LEP-I studies where a  +1\% $M_Z$ mass shift was also 
observed from line shape data. The left side of the ``data" distribution ({\it i.e.}
from the simulation including ISR and BS effects)
reproduces the ``ideal" line shape and, when fitted, gives an accuracy for the
$\Gamma_{Z'}$ measurement of about 2.5\%. In our fit we take into account an uncertainty
in the beam energy of $\delta E/E=10^{-4}$, 
as well as a statistical error based on the assumption that
we fit the production rate at 20 different equally spaced collider energies
centered around the $Z'$ mass
and have statistics corresponding to 10 fb$^{-1}$ of integrated
luminosity at each energy.

On the other hand, if we assume the same relative precision for inferring
the ``ideal" line shape from the ``data" as was obtained at  LEP-I
(this requires the same quality of  higher order corrections and event generators)
then the uncertainty in the determination of $M_{Z'}$
and $\Gamma_{Z'}$ would come mainly from the beam energy spread and 
statistical error.
In this case, a fit to the  ``ideal" data gives roughly a $0.2\%$
uncertainty for the measurement of  $\Gamma_{Z'}$ 
and a  $0.01\%$ uncertainty for the measurement of  $M_{Z'}$.

\begin{figure}
\begin{center}
\includegraphics[scale=1]{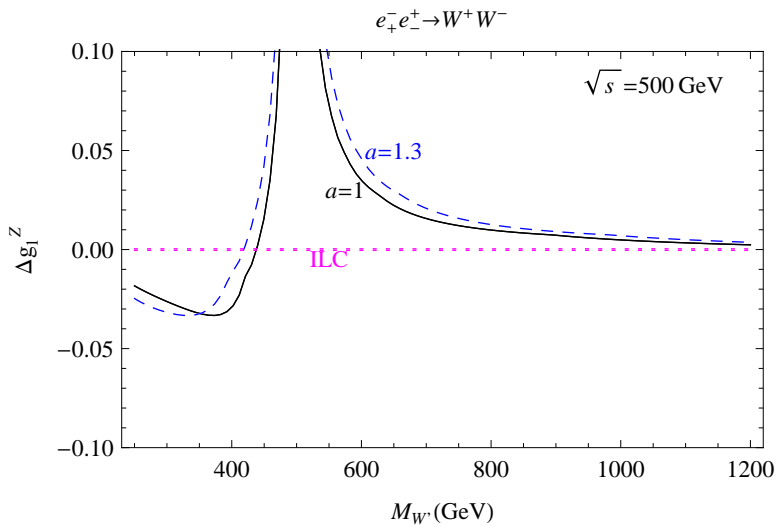}
\end{center}
\caption{\label{fig:ILC} The value of the effective $\Delta g_1^Z$  in the triangular 
Higgsless model, as discussed in section \protect\ref{sec:LEP}, for a center-of-mass
energy $\sqrt{s}=500$ GeV, as a function of $M_{W'}$ for various values of $a$
and ideal delocalization ($S_0 \approx 0)$. 
The value in both the three-site model and the minimal four-site model is
obtained by setting $a=1$. The anticipated sensitivity of the ILC to the value of $\Delta g_1^Z$ is given, approximately,
by the thickness of the dashed line labeled "ILC".}
\end{figure}

Even if the mass of the $Z'$ is higher than the energy of the International Linear
Collider, the triangular moose model can still be probed in a manner analogous to
the LEP-II  limits on the triangular moose model described in section \ref{sec:LEP}.  
The ILC will be sensitive to deviations of
$\Delta g_1^Z$ as small as $3.8\times10^{-4}$ \cite{Abe:2001nq}.  
Using the method described in
section \ref{sec:LEP}, we can estimate the effective $\Delta g_1^Z$ for this
model at the ILC.  We provide an illustrative example in Figure \ref{fig:ILC} for
$e_+^-e_-^+\rightarrow W^+W^-$ where the $e^-$ is right polarized and the $e^+$ is
left polarized.  We see that the effect of  ideally delocalized HLS-type
models will be observable at the ILC for the entire range of masses.  Furthermore,
if the mass of the $W'$ is known from the LHC with sufficient precision, we may be able to
determine the value of the parameter $a$ from this and other polarization channels.

\section{Conclusions}

We have considered how well the three-site model performs as a general representative 
of Higgsless models, and have studied several modifications which have the potential to improve upon its performance.  Our comparisons have employed sum rules \cite{Csaki:2003dt, SekharChivukula:2008mj} relating the masses and couplings of the gauge field KK modes, because these identities enable us to quantify how well a given theory unitarizes the scattering of electroweak gauge bosons at a given energy scale.  After comparing the three-site model to a pair of 5d continuum $SU(2) \times SU(2)$ models in flat and warped space, we have further analyzed two deconstructed extensions of the three-site model: a longer open linear moose with an additional $U(1)$ group and an  HLS or BESS ring model with three sites and three links.

We find that the tendency of the sum rules to be saturated by contributions from the lowest-lying KK resonances suggests a way to quantify the extent to which a highly-deconstructed theory like the three-site model can accurately describe the low-energy physics. Specifically,  the following quantity
\begin{equation}
 \hat{a} \equiv \frac43 (1 - g_{\pi\pi\pi\pi}) \approx \frac{v^2 g_{Z' WW}^2 M_{Z'}^2}{M_W^4}~,\nonumber
 \end{equation}
where the $Z'$ is the lightest neutral $KK$ boson,
provides a useful measure of the ability of  deconstructed models to approximate the behavior of  continuum theories.  In continuum theories, $\hat a = 4/3$ and the multi-pion coupling $g_{\pi\pi\pi\pi}$ vanishes; in the three-site and other linear deconstructed models we study, $\hat a = 1$ and $g_{\pi\pi\pi\pi} = \frac14 $; whereas, in the triangular ring model, $\hat{a}$ is essentially a free parameter, allowing interpolation between these cases. We have 
demonstrated that $W_LW_L$ scattering in the triangular ring model can, 
if the hidden local symmetry parameter $a$  is
chosen to mimic $\rho$-meson dominance of $\pi\pi$ scattering in QCD
 \cite{Harada:2003jx,Harada:2006di},  very closely approximate scattering in the continuum models.

Our study of hadron and lepton collider phenomenology of the extended models confirms that the three-site model will be a useful benchmark  against which to perform the first comparisons of new collider data with a whole class of Higgsless 5d and deconstructed theories.  For example, most of the gauge-boson and fermion properties and couplings characteristic of the three-site model are unaltered or minimally affected by the addition of an additional $U(1)$ group.  
In particular, the charged electroweak bosons of the four-site and three-site models are identical, while the pair of degenerate $Z'$ and $Z''$ bosons for $u^2\tilde{q}^2 [{f'/f}]^2=2$ (see Section III) jointly couple to $W$-boson pairs like the single $Z'$ boson of the three-site model.  We therefore conclude that vector-boson scattering processes among the lightest KK gauge bosons in a continuum $SU(2)^2 \times U(1)$ model with a flat extra dimension will be well-described by the three-site model.  Since the three-site model also provides an effective low-energy description of the $SU(2)\times SU(2)$ continuum model, it follows that the lightest KK gauge modes of the  $SU(2)^2 \times U(1)$ and $SU(2)^2$ continuum models will also appear phenomenologically similar at low energies.  As a second example, in the triangular moose model,  the signals of the $W'$ boson or the contributions of the $Z'$ boson to $WW$ scattering or $Z'W$ associated production will have a similar form as in the three site model; the signal strength will be proportional to $\hat{a}$, which could reduce the luminosity required for discovery.   Performing an actual measurement of the value of $\hat{a}$ will likely require the precision of a high-energy lepton collider.

\vspace{1cm}

\begin{acknowledgments}

This work was supported in part by the US National Science Foundation under
grants  PHY-0354226 and PHY-0551164. RSC and EHS gratefully 
acknowledge the hospitality and support of the Kavli Institute for Theoretical Physics (U.C. Santa Barbara), the Kavli Institute for Theoretical Physics - China  (C.A.S., Beijing), and the Aspen Center for Physics 
where some of this work was completed. HJH is supported in part by the NSFC 
under grants 10625522 and 10635030. M.K. is supported by Los Alamos National Laboratory under DE-AC52-06NA25396 and by a LANL Director's Fellowship.  M.T. is supported in part by the JSPS Grant-in-Aid for Scientific
Research No.20540263, and by Nagoya University Global COE Program,
Quest for Fundamental Principles in the Universe.   NDC would like to thank Ken Hsieh for helpful discussions.

\end{acknowledgments}

\appendix


\section{The gauge sector of the four-site model}

This appendix gives a fuller account of the gauge sector of the four-site model. Because the $SU(2) \times SU(2)$ gauge sector of this model is identical to that of the three-site model \cite{SekharChivukula:2006cg}, the charged-gauge boson mass-squared matrix is unaltered.   The $W$ and $W'$ boson masses and wavefunctions and the expression for $G_F$ in terms of the model parameters ($g$, $v$, $x$) are the same as in Ref. \cite{SekharChivukula:2006cg}.  For example, the $W$ mass and wavefunction have the form,
\begin{equation}
  M_W^2 = \frac{g^2 v^2}{4}\left[1-\frac{x^2}{4}+O(x^6) \right]\,, \qquad M_{W'}^2 = \tilde{g}^2 v^2 \left[ 1 + \frac{x^2}{4} + \frac{x^4}{16} + O(x^6) \right]\,,
\label{eq:M_W2}
\end{equation}
\begin{eqnarray}
\label{eq:Wwave}
  W^\mu &=& v_W^0 W_0^\mu + v_W^1 W_1^\mu \,,\qquad W'^\mu = -v_W^1 W_0^\mu + v_W^0 W_1^\mu \,,\\
 v_W^0 &=& 1-\frac{x^2}{8}+O(x^4)\,, \nonumber \\
v_W^1 &=& \frac{x}{2}+\frac{x^3}{16}+O(x^5)\, . \nonumber
\end{eqnarray}

The mass-squared matrix for the neutral gauge bosons, on the other hand, is enlarged by the addition of the second $U(1)$ group and takes the form\footnote{For clarity here, and in discussing the photon below, we allow for arbitrary $q$ and $\tilde{q}$; elsewhere we specialize
to the case $q=\tilde{q}$ and display explicitly the dependence on $\tilde{q}$.}
\begin{equation}
\frac{{\tilde g}^2 v^2}{2}
\left(
\begin{array}{cccc}
\ \ x^2\ \ &\ \ -x\ \ &0&0\\
\ \ -x\ \ &\ \ 2\ \ &\ \ -xt\ \ &0\\
0&\ \ -xt\ \ &x^2t^2 \left(1 + q^2 \left[\dfrac{f'}{f}\right]^2\right)&-xtu\,  q \tilde{q} \left[\dfrac{f'}{f}\right]^2\\
0&0&-xtu\, q \tilde{q} \left[\dfrac{f'}{f}\right]^2&u^2\, \tilde{q}^2 \left[\dfrac{f'}{f}\right]^2 
\end{array}
\right).
\label{eq:mass_matrix}
\end{equation}
The matrix has a zero eigenvalue, corresponding to the massless photon, with an eigenstate which may be written
\begin{equation}
  A^\mu = v_\gamma^0 W_0^\mu + v_\gamma^1 W_1^\mu 
          + v_\gamma^2 B_2^\mu 
          + v_\gamma^3 B_3^\mu
= \frac{e}{g} W_0^\mu + \frac{e}{\tilde g} W_1^\mu 
          + \frac{e}{g'} B_2^\mu 
          + \frac{e}{{\tilde g}'} B_3^\mu, 
\label{eq:gamma_wave}
\end{equation}
where the electric charge $e$ satisfies
\begin{equation}
 \frac{1}{e^2} = \frac{1}{g^2} + \frac{1}{{\tilde g}^2}
             + \frac{1}{g'^2} + \frac{q^2}{ \tilde{q}^2\tilde{g}'^2}
 = \frac{1}{g^2 s^2}\left[1+s^2\left(1+\frac{1}{u^2}\right)x^2 \right].
\end{equation}
where the last equality holds for $q = \tilde{q}$.
The light neutral gauge boson, which we associate with the 
$Z$, has a mass
\begin{equation}
 M_Z^2 = \frac{g^2 v^2}{4}\left[(1+t^2)  
 - \left\{\frac{(1-t^2)^2}{4}+\frac{t^4}{u^2} \right\}x^2 + O(x^4)\right],
\end{equation}
with a corresponding eigenvector 
\begin{eqnarray}
  Z^\mu &=& v_Z^0 W_0^\mu + v_Z^1 W_1^\mu 
          + v_Z^2 B_2^\mu 
          + v_Z^3 B_3^\mu, \nonumber \\
 v_Z^0 &=& c+c^3\left\{-\frac{1+2t^2-3t^4}{8}+\frac{t^4}{2u^2}\right\}x^2 
 + \cdots,\label{eq:vZ0}\\
 v_Z^1 &=& \frac{c(1-t^2)}{2}x + c^3\left\{\frac{(1-t^2)^3}{16} 
 + \frac{(3+t^2)t^4}{4u^2}\right\}x^3 + \cdots,\label{eq:vZ1}\\
 v_Z^2 &=& -s + sc^2\left\{-\frac{3-2t^2-t^4}{8} + 
 \frac{(2+t^2)t^2}{2u^2}\right\}x^2 + \cdots,\label{eq:vZ2}\\
 v_Z^3 &=& -\frac{s^2}{cu}x + \frac{cs^2}{8\tilde{q}^2 u^3}\left(\frac{f}{f'}\right)^2
 \left\{
 -\frac{4}{c^4}+\tilde{q}^2\left(\frac{f'}{f}\right)^2\left(-3u^2+(2t^2+t^4)(4+u^2)\right)
 \right\}x^3 + \cdots.\label{eq:vZ3}
\end{eqnarray}
In the limit as $u\to \infty$, the expressions describing the electric charge and the $Z$ boson recover the values of the three-site model.   

The neutral gauge sector also includes two heavy mass eigenstates instead of the single $Z'$ boson of the three-site model.  The form of the masses and wavefunctions depends significantly on whether the value of $u^2\tilde{q}^2 \left[{f'/f}\right]^2$ is close to 2 or quite different in value from 2.

\subsection{The degenerate case: $u^2 \tilde{q}^2 \left[\dfrac{f'}{f}\right]^2= 2$}

First, we consider the special case $u^2 = 2$.  Finishing the diagonalization of the mass-squared matrix (\ref{eq:mass_matrix}) yields the following masses and eigenvectors of the $Z'$ and $Z''$ bosons:
\begin{eqnarray}
 M^2_{Z' (Z'')} & =& \tilde{g}^2 v^2 
 \left[1 + \frac{1}{8u^2} \left( u^2 + t^2(4+u^2) \mp w \right)x^2 + O(x^4)  \right] ,\\
 Z'^\mu (Z''^\mu) &=& v_{Z' (Z'')}^0 W_0^\mu + v_{Z' (Z'')}^1 W_1^\mu 
          + v_{Z' (Z'')}^2 B_2^\mu 
          + v_{Z' (Z'')}^3 B_3^\mu, \label{eq:heavyZwave_usq2}
\end{eqnarray}
\begin{eqnarray}
 v_{Z' (Z'')}^0 &=& \frac{\sqrt{2}t^2u}{\sqrt{ w\{w\pm(u^2+t^2[-4+u^2])\}}} x + \cdots\,, \qquad  
 v_{Z' (Z'')}^2 = \frac{w\pm (u^2+t^2[4+u^2])}{8tu}\sqrt{\frac{\{w\pm(u^2+t^2[-4+u^2])\}}{2w}} x \cdots\,, \nonumber \\
 v_{Z' (Z'')}^1 &=& - \frac{2\sqrt{2} t^2 u} {\sqrt{ w\{w\pm(u^2+t^2[-4+u^2])\}}} + \cdots\,, \qquad
 v_{Z' (Z'')}^3 = \pm \sqrt{\frac{\{w\pm(u^2+t^2[-4+u^2])\}}{2w}}+ \cdots\,, \nonumber
\end{eqnarray}
where 
\begin{equation}
w \equiv \sqrt{u^4+2t^2u^2(-4+u^2)+t^4(4+u^2)^2}~.
\end{equation}
 In this case, the two heavy bosons are degenerate and their wavefunctions at site 1 and site 3 are both of $O(1)$ while those at site 0 and site 2 are of $O(x)$.

\subsection{The non-degenerate case: $u^2\tilde{q}^2 \left[\dfrac{f'}{f}\right]^2 - 2 = {{O}(1)}$ }

We now consider the alternative limit, where $u^2\tilde{q}^2 \left[{f'/f}\right]^2 \neq 2$.   Finishing the diagonalization of the mass-squared matrix (\ref{eq:mass_matrix}) yields the following mass and eigenvector for the $Z'$:
\begin{eqnarray}
 M_{Z'}^2 &=& \tilde{g}^2 v^2\left[1  
 + \frac{1}{4}\left(1+t^2\right)x^2 + O(x^4)\right],\\
  Z'^\mu &=& v_{Z'}^0 W_0^\mu + v_{Z'}^1 W_1^\mu 
          + v_{Z'}^2 B_2^\mu 
          + v_{Z'}^3 B_3^\mu, 
\label{eq:Z'wave}
\end{eqnarray}
\begin{eqnarray}
 v_{Z'}^0 &=& -\frac{x}{2}  + \cdots\,, \qquad\qquad\qquad\qquad
  v_{Z'}^2 = -\frac{t}{2}x + \cdots \,,\nonumber\\
 v_{Z'}^1 &=& 1+\frac{1}{8}\left(-1-t^2\right)x^2 + \cdots\,, \qquad
 v_{Z'}^3 = \frac{{f'}^2\tilde{q}^2 t^2 u}{2(2f^2-{f'}^2\tilde{q}^2 u^2)}x^2 
     + \cdots\,, \nonumber
\end{eqnarray}
and the following mass and eigenvector for the $Z''$:
\begin{eqnarray}
 M_{Z''}^2 &=&\frac{\tilde{g}^2 v^2}{2}\left[\frac{f'}{f}\right]^2\,\tilde{q}^2\left[u^2 
 + t^2x^2 + O(x^4)\right],\\
  Z''^\mu &=& v_{Z''}^0 W_0^\mu + v_{Z''}^1 W_1^\mu 
          + v_{Z''}^2 B_2^\mu 
          + v_{Z''}^3 B_3^\mu, 
\end{eqnarray}
\begin{eqnarray}
v_{Z''}^0 &=& -\frac{f^4 t^2}{\tilde{q}{f'}^2u^3(2f^2-\tilde{q}^2 {f'}^2 u^2)}x^3  + \cdots\,, 
     \qquad\qquad\qquad v_{Z''}^2 = \frac{t}{u}x 
     + \cdots\,, \nonumber\\
v_{Z''}^1 &=& \frac{f^2 t^2}{u(2f^2-\tilde{q}^2{f'}^2u^2)}x^2 
    + \cdots \,,\qquad\qquad\qquad\qquad\quad
     v_{Z''}^3 = -1 + \frac{t^2}{2u^2}x^2 + \cdots\, . \nonumber
\end{eqnarray}
Clearly, the $Z'$ boson is strongly concentrated at site 1 and the $Z''$ boson, at site 3; the ratio of masses is $\left(M_{Z''}/M_{Z'}\right)^2 = \frac{u^2 \tilde{q}^2}{2}\left[\frac{f'}{f}\right]^2[1+O(x^2)] $.  In the $u \rightarrow \infty$ limit, we recover the three-site
 model: the $Z'$ mass and wavefunction revert to the three-site form, while the $Z''$ becomes infinitely massive and localized at site 3.

\section{The fermion sector of the four-site model}

The couplings of the $Z$ to the light fermion mass eigenstates include contributions from all four sites.
The left-handed coupling of the light $Z$-boson to quark fields may be written
\begin{equation}
{\cal L}_{ZL} \propto
Z_\mu\left[ g\,v_Z^0 (\bar{Q}_{L0} \frac{\tau^3}{2} \gamma^\mu  Q_{L0}) + 
\tilde{g}\,v_Z^1 (\bar{Q}_{L1} \frac{\tau^3}{2} \gamma^\mu Q_{L1})+
\frac{g'}{6}\,v_Z^2 \bar{Q}_{L0} \gamma^\mu Q_{L0}
+\frac{\tilde{g}'}{6}\,v_Z^3\bar{Q}_{L1} \gamma^\mu Q_{L1}\right]~,
\label{eq:leftZlag}
\end{equation}
where the first two terms give rise to the left-handed $T_3$ coupling and
the last two terms give rise to the left-handed hypercharge
coupling. The expression for leptons is similar, replacing hypercharge
$\frac16$ with $-\frac12$. Similarly, the right-handed coupling of the $Z$ to quarks fields is
\begin{equation}
{\cal L}_{ZR} \propto
Z_\mu\left[ \tilde{g}\,v_Z^1 (\bar{Q}_{R1} \frac{\tau^3}{2} \gamma^\mu  Q_{R1}) + 
\frac{\tilde{g}'}{6}\,v_Z^3 (\bar{Q}_{R1} \gamma^\mu Q_{R1})+
g'\,v_Z^2 \left(\frac23 \bar{u}_{R2} \gamma^\mu u_{R2}
-\frac13 \bar{d}_{R2} \gamma^\mu d_{R2}\right)\right]~,
\end{equation}
where the last three terms arise from the hypercharge. For leptons, $\frac16 \to -\frac12$
in the second term, $\frac23 \to 0$ in the third term (for neutrinos), and $-\frac13 \to -1$ in
the fourth term for the charged leptons. For quarks, this expression may
be more conveniently rewritten as
\begin{eqnarray}
{\cal L}_{ZR} &\propto&
Z_\mu\left[ (\tilde{g}\,v_Z^1-\tilde{g}' v_Z^3) (\bar{Q}_{R1} \frac{\tau^3}{2} \gamma^\mu  Q_{R1}) + 
\tilde{g}'\,v_Z^3 \left(\frac23 \bar{u}_{R1}\gamma^\mu u_{R1}
- \frac13 \bar{d}_{R1} \gamma^\mu d_{R1}\right) \right.\nonumber\\
&& \hspace{57mm}\left. +g'\,v_Z^2 \left(\frac23  \bar{u}_{R2}\gamma^\mu u_{R2}
- \frac13 \bar{d}_{R2} \gamma^\mu d_{R2}\right) \right],
\label{eq:rightZlag}
\end{eqnarray}
where the last four terms yield the $Z$-couplings to the conventionally-defined right-handed
hypercharge of the quarks, while the first can give rise to a new right-handed $T_3$ coupling.

From these expressions, we may derive the $T_3$ and hypercharge couplings between the fermions and the $Z$.  For ideally delocalized light fermions, 
we find the left-handed coupling to $T_3$ to be
\begin{eqnarray}
 g_{3L}^{Zqq} &=& g(b_L^0)^2(v_Z^0) + \tilde{g}(b_L^1)^2(v_Z^1)\nonumber\\
 &=& gc \left(1+\left\{-\frac{c^2(3+6t^2-t^4)}{8}
 +\frac{c^2t^4}{2u^2}\right\}x^2+O(x^4)\right)\nonumber\\
 &=& \frac{eM_W}{M_Z\sqrt{1-\frac{M_W^2}{M_Z^2}}}
 \left[1+O(x^4)\right].\label{eq:g3LZqq-2}
\end{eqnarray}
The expression in terms of $s$ is explicitly dependent on $u$, but when written in terms of the physical quantities $e$, $M_Z$, and $M_W$, the form of $g_{3L}^{Zqq}$ is nearly of Standard Model form for arbitrary values of $u$, as we expect for ideal delocalization.  Since the right-handed light fermion eigenvectors are localized entirely at site 2, there are no right-handed couplings of the light fermions to $T_3$.  The left- and right-handed couplings of the top-quark to $T_3$ are
\begin{equation}
 g_{3L}^{Ztt} = g_{3L}^{Zqq}\left(1+\frac{\varepsilon_{tR}^2(2+\varepsilon_{tR}^2)}
{4c^2(1+\varepsilon_{tR}^2)^2}x^2+O(x^4)\right)\,, \qquad\qquad  g_{3R}^{Ztt} = \frac{g\varepsilon_{tR}^2} {2 c (1+\varepsilon_{tR}^2)}\left[1 + O(x^2)\right]~,
\label{eq:g3LZtt-2}
\end{equation}
as in the three-site model; similar expressions hold for the bottom quark, 
with $\varepsilon_{tR}\rightarrow \varepsilon_{bR}$.  

Again, for ideally delocalized light fermions,  the left-handed coupling to $Y$ is
\begin{eqnarray}
 g_{YL}^{Zqq} &=& g'(b_L^0)^2(v_Z^2) + {\tilde g}'(b_L^1)^2(v_Z^3)\nonumber\\
 &=& -g's \left(1+\left\{\frac{c^2(3-2t^2-t^4)}{8}
 -\frac{c^2t^2(2+t^2)}{2u^2}\right\}x^2+O(x^4)\right)\nonumber\\
 &=& -\frac{eM_Z}{M_W}\sqrt{1-\frac{M_W^2}{M_Z^2}}
 \left[1+O(x^4)\right]~,
 \label{eq:gYLZqq-2}
\end{eqnarray}
which is nearly of Standard Model form when written in terms of physical quantities.  The hypercharge couplings of the right-handed light fermions and the left-handed top quark to the $Z$ are also of this form to $O(x^2)$:
\begin{equation}
  g_{YR}^{Zqq} =  g_{YL}^{Zqq}\left[1+O(x^4)\right]\,,   \qquad\qquad  g_{YL}^{Ztt} = g_{YL}^{Zqq}\left[1+O(x^4)\right].
\end{equation}
In the three-site model, the hypercharge coupling of the right-handed top quark to the $Z$ is also of this form because the hypercharge is entirely localized at site 2.  However, in the four-site model, the hypercharge coupling of the top quark comes from both site 2 and site 3 and the $Z$-boson wavefunction is not perfectly flat on those sites; the component at the second site $g' v^2_Z$ differs at $O(x^2)$ from that at the third site  $\tilde{g}' v^3_Z$, as may be seen by comparing Eqs. (\ref{eq:vZ2}) and (\ref{eq:vZ3}).  The result is that the top-quark's right-handed hypercharge coupling differs from that of the light quarks:
\begin{equation}
  g_{YR}^{Ztt} = {\tilde g}'(t_R^1)^2(v_Z^3)+g'(t_R^2)^2(v_Z^2) = g_{YR}^{Zqq}\left[1+\frac{1}{2c^2u^2}
\frac{\varepsilon_{tR}^2}{(1+\varepsilon_{tR}^2)}x^2+O(x^4)\right]~,
\label{eq:gYRZtt-2}
\end{equation}
at $O(x^2)$.

\section{The Gauge Sector of the Triangular Three-Site Model}

The charged-boson mass matrix is given by
\begin{equation}
{\cal M}^2_W = 
\frac{\tilde{g}^2 v^2}{4}
\begin{pmatrix}
x^2(1+a) & -2xa \\
-2xa & 4a
\end{pmatrix}~,
\end{equation}
where $x=g/\tilde{g} \ll 1$. Expanding in $x$, we find
the masses
\begin{equation}
M^2_W = \frac{g^2 v^2}{4} \left(1-\frac{x^2}{4} + \frac{(a-1) x^4}{16 a} + \cdots\right)\,, \qquad \qquad
M^2_{W'}  = \tilde{g}^2 v^2 a \left(1+\frac{x^2}{4} + \frac{x^4}{16 a} + \cdots\right)~,
\end{equation}
from which we may derive the relationship
\begin{equation}
x^2 = 4 a \left(\frac{M_W}{M_{W'}}\right)^2 + 8 a^2 \left(\frac{M_W}{M_{W'}}\right)^4
+ 4 a^2(5a+2) \left(\frac{M_W}{M_{W'}}\right)^6+\cdots~.
\end{equation}
The light and heavy charged gauge boson eigenstates are given, respectively, by
\begin{eqnarray}
W^\mu = v_{W}^0 W_0^\mu &+& v_{W}^1 W_1^\mu, \label{eq:wvec}\\
v_{W}^0 &=& 1-\frac{x^2}{8} + \frac{(3a-8)x^4}{128 a} + \cdots\,, \qquad\qquad 
 v_{W}^1 = \frac{x}{2} - \frac{(a-2)x^3}{16 a} + \frac{(3a^2-20a+8)x^5}{256 a^2} + \cdots\,, \nonumber\\
W'^\mu = v_{W'}^0 W_0^\mu &+& v_{W'}^1 W_1^\mu,  \label{eq:wpvec}\\
    v_{W'}^0 &=& - \frac{x}{2} - \frac{2-a}{16 a} x^3 + \cdots \,, \qquad\qquad\qquad\qquad
    v_{W'}^1 = 1 - \frac{x^2}{8} + \cdots .\nonumber 
\end{eqnarray}

The neutral-boson mass matrix is given by
\begin{equation}
{\cal M}^2_Z = 
\frac{\tilde{g}^2 v^2}{4} 
\begin{pmatrix}
x^2(1+a) & -2xa & -t x^2 (1-a) \\
-2xa & 4a & -2txa\\
-t x^2 (1-a) & -2txa & t^2 x^2 (1+a)
\end{pmatrix}~,
\end{equation}
where $t\equiv g'/g = s/c$ and $s^2 + c^2 = 1$. This matrix has a zero eigenvalue, corresponding to the photon,
which has the following eigenvector:
\begin{equation}
A^\mu = \frac{e}{g} W_0^\mu + \frac{e}{\tilde{g}} W_1^\mu + \frac{e}{g'} B^\mu \,, \qquad\qquad\qquad\qquad   \frac{1}{e^2} = \frac{1}{g^2} + \frac{1}{\tilde{g}^2} + \frac{1}{{g'}^2}\,,
\end{equation}
where $B^\mu$ is the $U(1)$ gauge boson associated with site 2.  The light neutral gauge boson, which we associate with the $Z$, has a mass
\begin{equation}
 M^2_Z  = \frac{g^2 v^2}{4 c^2}
\left[ 1- \frac{x^2}{4} \frac{(c^2-s^2)^2}{c^2} + \frac{(a-1)(s^2-c^2)^2 x^4}{16 c^4 a} +\cdots\right]\,,
\end{equation}
with a corresponding eigenvector
\begin{eqnarray}
Z^\mu &=& v_{Z}^0 W_0^\mu + v_{Z}^1 W_1^\mu + v_{Z}^2 B^\mu\,, \label{eq:triang-z}  \\
 && v_{Z}^0 =  c - \frac{c^3 (1+2t^2-3t^4)}{8}x^2 + \cdots ,\nonumber \\
 && v_{Z}^1 =  \frac{c(1-t^2)}{2}x + \frac{c^3(1-t^2)\left(\, 2(1+t^2)^2-a(1+6t^2+t^4)\, \right)}{16a} x^3 + \cdots ,\nonumber \\
 && v_{Z}^2 =  -s - \frac{s c^2 (3-2t^2-t^4)}{8} x^2 \cdots . \nonumber 
\end{eqnarray}
The heavy neutral $Z'$ boson has a mass
\begin{equation}
 M^2_{Z'}  = \tilde{g}^2 v^2 a \left[1+ \frac{x^2}{4 c^2} + \frac{x^4(1-t^2)^2}{16 a} + \cdots\right]~,
\end{equation}
and eigenvector
\begin{eqnarray}
 \label{eq:triang-zp}
{Z'}^\mu &=& v_{Z'}^0 W_0^\mu + v_{Z'}^1 W_1^\mu + v_{Z'}^2 B^\mu ,  \\
  v_{Z'}^0 &=&  -\frac{x}{2} + \frac{-2a(1-t^2)+a^2(1+t^2)}{16a^2}x^3 + \cdots ,\nonumber \\
 v_{Z'}^1 &=&  1 - \frac{(1+t^2)}{8}x^2 + \cdots ,\nonumber \\
  v_{Z'}^2 &=&   -\frac{t}{2}x + \frac{t (\,2(1-t^2) + a (1+t^2)\,)}{16a} x^3 + \cdots .\nonumber 
\end{eqnarray}
Note that all the expressions for quantities related to the gauge sector reduce to their counterparts in the three-site model if we set $a=1$.


\end{document}